\documentclass[prc,twocolumn,epsfig,nofootinbib,floatfix]{revtex4}
\usepackage{graphics}
\usepackage{epsfig}
\usepackage{amsfonts}
\usepackage{amsmath}
\usepackage{bm}

\usepackage{color}

\begin{document}
\title{GENERAL TREATMENT OF VORTICAL, TOROIDAL, AND COMPRESSION MODES}
\author{J. Kvasil$^1$,V.O. Nesterenko$^2$, W. Kleinig$^{2,3}$,
P.-G. Reinhard$^4$, and P. Vesely$^{1,5}$}
\affiliation{$^1$
Institute of Particle and Nuclear Physics, Charles
University, CZ-18000, Praha 8, Czech Republic}
\email{kvasil@ipnp.troja.mff.cuni.cz}
\affiliation{$^2$
Laboratory of Theoretical Physics,
Joint Institute for Nuclear Research, Dubna, Moscow
region, 141980, Russia}
\email{nester@theor.jinr.ru}
\affiliation{$^3$
Technische Universit\"at Dresden, Inst. f\"ur Analysis,D-01062,
Dresden, Germany}
\affiliation{$^4$ Institut f\"ur Theoretische Physik II,
Universit\"at Erlangen, D-91058, Erlangen, Germany}
\affiliation{$^5$
Department of Physics, P.O. Box 35 (YFL)
FI-40014, University of Jyv\"askyl\"a, Finland}

\date{\today}

\begin{abstract}
The multipole vortical, toroidal, and compression modes are analyzed. Following
the vorticity concept of Ravenhall and Wambach,
the vortical operator is derived and related in a simple way  to the toroidal
and compression operators. The strength functions and velocity fields of the
modes are analyzed in $^{208}$Pb within the random-phase-approximation using
the Skyrme force SLy6. Both convection and magnetization nuclear currents are
taken into account. It is shown that the isoscalar (isovector) vortical and
toroidal modes are dominated by the convection (magnetization) nuclear current
while the compression mode is fully convective. The relation between the above
concept of the vorticity  to the hydrodynamical vorticity  is briefly
discussed.
\end{abstract}


\maketitle

\section{Introduction}
\label{sec:introduction}

An irrotational character of nuclear flow is a basic assumption in collective
nuclear dynamics \cite{Bo74,Forest_Walecka_66,Hei_81,Ri80}, which manifests
itself in numerous examples of low-energy excitations and giant resonances
(GR).  At the same time, nuclear motion can also carry a vorticity, i.e. a
deviation from irrotational flow \cite{Bertsch_NPA_75,Se83,Ra87,Ca99}. In
hydrodynamics (HD), the vorticity is defined as a curl of the velocity field
\cite{Lan87B}. Instead, the nuclear theory deals with nuclear currents rather than
velocities and so here the vorticity is often defined through the
$j_{\lambda \lambda+1}(r)$ component of the multipole
decomposition of the transverse nuclear current \cite{Ra87}.
The component $j_{\lambda \lambda+1}(r)$ is treated as unrestricted by the
continuity equation (containing the current divergence $\vec{\nabla} \cdot
\vec{j}$) and so is believed to be of a vortical character. In this case, the
vorticity and charge transition density represent two independent parts of  the
charge-current distribution. This consideration reminds the previous result
\cite{Hei_PRC_82} where the current component $j_{\lambda \lambda-1}(r)$ is
proposed to be constrained by the continuity equation and thus determined by
the charge distribution while the component $j_{\lambda \lambda+1}(r)$ is
treated as independent.

Both definitions of the vorticity, from HD and Ref. \cite{Ra87}, are
widely used in the literature: the former in the nuclear fluid-dynamical
models (see, e.g. \cite{Mi06} and references therein) and the latter in the
microscopic studies, see e.g. \cite{Ra87,Ca99}.
These two definitions are assumed to be closely related \cite{Ra87},
though they are different observables by construction. Actually, they represent
different aspects of the nuclear vorticity.
In this paper, we will concentrate on the  $j_{\lambda \lambda+1}(r)$-based
vorticity \cite{Ra87}. The comparison with the HD case will be also done.

The most remarkable manifestation of vorticity is the electric dipole toroidal
mode (TM) \cite{Semenko_81,Dub_75_83,Cl01} intensively explored during
the last decades, see e.g. the review \cite{Pa07} and references therein.
This exotic mode is treated as a vortical collective motion of the toroidal type.
The TM operator is the
second-order correction to the leading E1 operator in the long-wave approximation.
Another kind of the second-order E1 flow is represented by the anisotropic
compression mode (CM), often called the isoscalar dipole GR
\cite{Harakeh_PRL_77,Young_PRL_77,Morsch_PRL_80,Stringari_PLB_82}.
The mode is viewed as a compression wave in a definite direction and so is
related to the nuclear incompressibility
\cite{Harakeh_PRL_77,Stringari_PLB_82}. The isoscalar (T=0) TM and CM were
observed in $(\alpha, \alpha')$-reaction as broad low-energy (TM dominated) and
high-energy (CM dominated) electric dipole distributions
\cite{Cl01,Morsch_PRL_80,Morsch_PRC_83,Adams_PRC_86,Davis_PRC_97,
Yo04,Uchida_PLB_03,Uchida_PRC_04}. The TM was also investigated in the region
of the pygmy resonance in $^{208}$Pb in a nuclear fluorescence experiment
\cite{Ry02}. Perhaps, the TM can be discriminated in the transverse $(e,e')$
form-factors \cite{Richter_NPA_04}.

The TM and CM were examined in various models, including the fluid-dynamical
and sum-rule approaches \cite{Mi06,Semenko_81,Stringari_PLB_82,Bast93}, the
method of Wigner function moments \cite{Balbutsev_JPG_88,Balbutsev_EPL_94}, the
random-phase-approximation (RPA) \cite{Kv03} and multi-phonon \cite{Ry02}
methods with phenomenological single-particle potentials. More refined RPA
studies within the self-consistent mean-field approaches were also performed,
relativistic ones \cite{Vr00,Vr02} and those based on Skyrme forces
\cite{Co00,Papa_PLB_04}, for a review see \cite{Pa07}. A direct relation
between the current-dependent TM and density-dependent CM operators was
established in \cite{Semenko_81,Kv03} and both modes were shown to be mixed.
Most of the studies reproduce the observed bimodal (low-energy TM and
high-energy CM) distribution. However, theoretical models generally
overestimate the CM peak energy by $\sim$4 MeV and underestimates the TM one by
1-2 MeV. Besides, they yield a much broader TM distribution \cite{Pa07}.

A special effort was devoted to the nuclear vorticity as such
\cite{Bertsch_NPA_75,Se83,Ra87,Ca99}.
The possibility to measure the vorticity in $(e,e')$ experiments was
discussed \cite{Richter_NPA_04,Wu87,Ca99}.

Despite these thorough studies, some principle points concerning the
vorticity and related modes deserve further inspection:
i) There is an
essential difference in modeling the vortical mode (VM) and their TM
and CM counterparts.  The TM and CM are usually deduced as second
order terms in a low-momentum expansion of the electric multipole
transition operators \cite{Semenko_81,Dub_75_83}.  To the best of our
knowledge, an analogous way to the VM operator has yet to be
developed. This would help to establish a formal relation between VM
and TM/CM. ii) Most of the previous studies
(with exception of \cite{Ra87,Kv03,Ca99})
employ only the convection part $j_c$ of the nuclear
current and skip its magnetization (spin) part $j_{m}$, though the
latter can also produce a vorticity. The role of $j_m$ in VM and
other modes has yet to be clarified.
iii) Mainly the T=0 channel of the modes were
discussed, although their T=1 counterpart is also interesting and
deserves a closer look.
iv) The relation between two definition of the vorticity,
from HD \cite{Lan87B} and
$j_{\lambda\lambda+1}(r)$ current component \cite{Ra87}, has
yet to be clarified.

The aim of the present study is to explore the open problems listed above.
First of all, the vortical operator unconstrained by the charge conservation
is derived following the ideas \cite{Ra87}. The operator has a simple
relation with its TM/CM counterparts and is also deduced  as a second-order
term in a low-momentum (long
wavelength) expansion of the dominant electric operator.  Further, the
difference in the vorticity criteria from the HD \cite{Lan87B}
and Ref. \cite{Ra87} is inspected. In the numerical
calculations, both T=0 and T=1 channels of VM, TM, and CM are analyzed
by using the full nuclear current $j=j_c+j_{m}$.  The dominant role of
$j_{m}$ and thus the spin vorticity in forming the isovector VM and
TM is worked out. Both single-particle and collective mechanisms of
the vorticity are discussed.

The numerical analysis is done within the {\it self-consistent}
separable random-phase-approximation (SRPA) approach based on the
factorized Skyrme residual interaction
\cite{nest_PRC_02,nest_PRC_06}. The systematic studies of electric
\cite{nest_PRC_02,nest_PRC_06,nest_IJMPE_07,nest_IJMPE_08,nest_PRC_08,Kvasil_IJMPE_09}
and magnetic \cite{Ve09,Nest_JPG_10,Nest_IJMPE_10} GR in spherical and
deformed nuclei have established this approach as  a reliable and
effective theoretical tool.

The paper is organized as follows.  In Sec. II, the nuclear vorticity
is discussed in context of the previous studies.  In Sec. III, the VM,
TM, and CM operators are derived on the same theoretical grounds,
following the prescription \cite{Ra87}. A simple relation between the
operators is established. The mode velocities are analyzed and different
criteria of the vorticity, from \cite{Ra87} and HD, are compared.
Sec. IV provides an outline of the calculation scheme within the Skyrme SRPA.
In Sec. V, the numerical results for the T=0 and T=1 VM, TM, and CM are
discussed. A summary is given in Sec. VI.
Appendix \ref{Sec:_constraint} justifies the procedure
of building the vortical operator.
Appendix \ref{Sec:_spurious} describes  the procedure for
extraction of the spurious center-of-mass corrections.
Appendix \ref{Sec:_density_oper} provides the explicit expressions
for the density and current operators.
Appendix \ref{Sec:_SRPA} sketches the basics of SRPA.

\section{Vorticity in terms of velocities and currents}

The HD nuclear models, including the famous liquid-drop
model, assume an irrotational character of the collective nuclear flow
\cite{Bo74,Forest_Walecka_66,Hei_81}
\begin{equation}\label{irrot_v}
\vec{\nabla} \times \vec{v} (\vec{r})=0
\end{equation}
where $\vec{v}(\vec{r})$ is the local velocity field.
The deviation from the irrotational flow is quantified by the HD vorticity
\begin{equation}\label{vort}
  \vec \varpi ({\vec r})= \vec{\nabla} \times {\vec v}({\vec r})
   \;.
\end{equation}
Unlike the HD models, nuclear theory prefers to deal
with currents ${\vec j}_{nuc}({\vec r})$ rather than
velocities ${\vec v}({\vec r})$.
However, Eqs. (\ref{irrot_v}) and (\ref{vort}) cannot be replaced by the
similar expressions for the nuclear current
since, as shown below, the curls of velocity and current have a
different structure. Moreover, $\vec{\nabla} \times \vec{j}_{nuc} (\vec{r})$
is the key part of the electrical multipole operator
$\hat{M}(E\lambda\mu,k)$
and, so, treating $\vec{\nabla} \times \vec{j}_{nuc} (\vec{r})$
as a vortical quantity would wrongly mean a fully vortical nature of any
$E\lambda\mu$ excitations, which contradicts, e.g., a predominantly
irrotational character of electric GR.

For using the HD definition of vorticity (\ref{vort}),
the quantum theory should express it
through the nuclear current. A common way is to define the
velocity field $\vec{v}_{\nu}(\vec{r})$ for the excitation mode $\nu$
through the current transition density $\delta \vec{j}_{\nu}(\vec{r})$ and
ground state density $\rho_0(\vec{r})$ \cite{Se83,Suzuki_Rowe_NPA},
\begin{equation}
   \delta \vec{j}_{\nu}(\vec{r}) = \rho_0 (\vec{r})\vec{v}_{\nu}(\vec{r}) \;,
  \label{vel_curr}
\end{equation}
which casts (\ref{vort}) into the form
\begin{equation}\label{vor_v_j}
\vec{\nabla} \times {\vec v}_{\nu}({\vec r})= \frac{\rho_0({\vec
r})\vec{\nabla} \times \delta{\vec j}_{\nu}({\vec r})
 - \vec{\nabla}\rho_0({\vec r})\times \delta{\vec j}_{\nu}({\vec r})}
{\rho_0^2({\vec r})} \; .
\end{equation}
This illuminates the difference between curls of the
velocity and current and thus shows that
$\vec{\nabla} \times \delta{\vec j}_{\nu}({\vec r})$
cannot be a measure of vorticity. The difference is comprised in
the gradient field $\vec{\nabla}\rho_0({\vec r})$ which is maximal at the
nuclear surface and minimal in the interior.

In \cite{Ra87}, a concept of nuclear vorticity, alternative to the HD one and fully
based on the nuclear current, was proposed. It aims
to find a  component of the nuclear current which is not
restricted by the continuity equation
\begin{equation}\label{ce_ctd}
\vec{\nabla} \cdot \delta\vec{j}_{\nu}(\vec{r}) = - ikc \delta\rho_{\nu} (\vec{r})
\end{equation}
with $k$ being the transfer momentum. By construction, this current component should
not contribute to the irrotational motion and
vanish in the divergence $\vec{\nabla} \cdot \delta\vec{j}_{\nu}(\vec{r})$.
So it may be naturally used for building the vortical quantities.

Since the present study follows similar lines, it is worth to
outline the concept \cite{Ra87} in more detail.  In spherical nuclei,
it exploits the multipole expansion of the nucleon and current
transition densities:
\begin{eqnarray}
\nonumber
\delta \rho_{fi}(\vec{r}) &=& \langle j_f m_f|\:\hat{\rho}(\vec{r}) \:| j_i m_i \rangle
\\ \label{td}
&=& \sum_{\lambda \mu} a^{fi}_{\lambda \mu}\:
\rho_{\lambda}(r) \: Y_{\lambda\mu} (\hat{\vec r}) \; ,
\\
\label{ctd}
\nonumber
\delta \vec{j}_{fi}(\vec{r})
&=&\langle j_f m_f| \: \hat{\vec{j}}_{\text nuc}(\vec{r}) \:| j_i m_i \rangle
\\
&=& -i\sum_{\lambda \mu}\sum_{L=\lambda \pm 1}
a^{fi}_{\lambda \mu} \:
j_{\lambda\:  L}(r) \:\vec{Y}_{\lambda L \mu}(\hat{\vec r}) \ ,
\end{eqnarray}
where
\begin{equation}\label{acoef}
a^{fi}_{\lambda \mu} = (-1)^{\mu} \frac{(j_i m_i \lambda -\mu|j_f
m_f)}{\sqrt{2j_f + 1}} \; ,
\end{equation}
$j_i, m_i$ ($j_f, m_f$) are spins and their projections for the initial $i$
(final $f$) state, $Y_{\lambda\mu} (\hat{\vec r})$ and $\vec{Y}_{\lambda L
\mu}(\hat{\vec r})$ are ordinary and vector spherical harmonics \cite{Va76}.
The $\rho_\lambda(r)$ and $j_{\lambda L}(r)$ are multipole components of
the transition density and current.
Using the above expansions and the quantity
\begin{equation}\label{std}
\delta \vec{S}_{fi}(\vec{r})
= \sum_{\lambda \mu} a^{fi}_{\lambda \mu}
\sqrt{\frac{\lambda + 1}{\lambda}}
kc \rho_{\lambda}(r) \:\vec{Y}_{\lambda \lambda \mu}(\hat{\vec r}) \ ,
\end{equation}
the unconstrained vortical transition density
\begin{eqnarray} \label{vtd}
\delta \vec{w}_{fi}(\vec{r}) &=&
\vec{\nabla} \times \delta \vec{j}_{fi}(\vec{r}) - \delta \vec{S}_{fi}(\vec{r})
\\
\nonumber
&=&
\sum_{\lambda \mu} a^{fi}_{\lambda \mu}
w_{\lambda\lambda}(r) \:\vec{Y}_{\lambda \lambda \mu}(\hat{\vec r})
\end{eqnarray}
is built \cite{Ra87}, where the vortical multipoles
\begin{equation}
w_{\lambda \lambda}(r) =
\sqrt{\frac{2 \lambda +1}{\lambda}}\left(\frac{d}{dr}
+ \frac{\lambda+2}{r}\right) \:j_{\lambda \: \lambda+1}(r)
\label{ctdm}
\end{equation}
are  determined by the radial current component
$j_{\lambda \:\lambda+1}(r)$. Finally,  the vorticity strength is given by
\begin{equation}
  \nu_{\lambda}
  =
  \int^{\infty}_{0} dr\,r^{\lambda +4}\:
  w_{\lambda \: \lambda}(r)\; .
\label{vs}
\end{equation}

The continuity equation (\ref{ce_ctd}) in terms of the $r^{\lambda}$ moments  relates
the current component $j_{\lambda\:\lambda-1}(r)$  to the
transition density but leaves  the component
$j_{\lambda\:\lambda+1}(r)$  untouched \cite{Ra87}.
So, just $j_{\lambda\:\lambda-1}(r)$ provides the charge-conservation
constraint and the quantity (\ref{std}) is constructed so as
to remove $j_{\lambda \:\lambda-1}(r)$ from
$\vec{\nabla} \times \delta \vec{j}_{fi}(\vec{r})$,
see more discussion in Appendix \ref{Sec:_constraint}.
The unconstrained vortical value
(\ref{ctdm}) includes only $j_{\lambda \:\lambda+1}(r)$.
Moreover, following \cite{Ra87}, the motion is treated as vortical if its
current involves $j_{\lambda \:\lambda+1}(r)$.


The formalism \cite{Ra87} treats the vorticity
without using an explicit vortical operator analogous to the
TM and CM ones. However, we need such operator for reasons of
comparison of vortical modes with TM and CM.
In the next section, we will develop the explicit vortical operator by
using the prescription \cite{Ra87} and relate this operator
with the toroidal and compression ones.
The subsequent discussion will show that this operator is
not truly vortical in the HD sense and
the presence of $j_{\lambda \:\lambda+1}(r)$ in the
current is not enough for the HD vorticity.

\section{VM, TM, and CM operators}

\subsection{Derivation of VM, TM and, CM operators}
\label{sec:derive}

The standard electrical multipole operator
 may be written in different forms \cite{BMv1}:
\begin{eqnarray}\label{ME_oper1}
\hat{M}(E\lambda\mu, k) &=&
-i \frac{(2 \lambda + 1)!!}{c k^{\lambda+1} (\lambda+1)}
\\
&&\cdot\int\!d^3r
\hat{\vec{j}}_{\text nuc}(\vec{r})\!\cdot\!
[\vec{\nabla}\!\times\!(\vec{r}
\!\times\!\vec{\nabla})j_{\lambda}(kr) Y_{\lambda \mu}(\hat{\vec r})]
\nonumber
\\ \label{ME_oper2}
&=& \frac{(2 \lambda + 1)!!}{c k^{\lambda+1}}
\sqrt{\frac{\lambda}{\lambda+1}}
\\
&& \cdot\int d^3r \: [\: j_{\lambda}(kr)\:
\vec{Y}_{\lambda \lambda \mu}(\hat{\vec r}) \:]
\cdot [ \vec{\nabla} \times \hat{\vec{j}}_{\text nuc}(\vec{r})]
\nonumber
\end{eqnarray}
where $j_{\lambda}(kr)$ is the spherical Bessel function.

The form (\ref{ME_oper2}) shows that $\vec{\nabla} \times
\hat{\vec{j}}_{nuc}(\vec{r})$ cannot be a measure of the vorticity since
otherwise $\hat{M}(E\lambda\mu, k)$ would indicate only vortical electric
excitations. At the same time, the form (\ref{ME_oper2})
suggests that the vortical operator may be built from $\hat{M}(Ek\lambda \mu)$
by replacing $\vec{\nabla} \times \hat{\vec{j}}_{nuc}(\vec{r})$ with  the truly
vortical quantity \cite{Ra87}
\begin{equation}
\vec{\nabla} \times \hat{\vec{j}}_{\text nuc}(\vec{r})
- \frac{i}{\lambda}kc \; [\vec{\nabla}\hat{\rho}(\vec{r}) \times \vec{r} ]
 \; .
\label{12}
\end{equation}
The density-dependent term in (\ref{12}) subtracts the charge conservation
constraint. Actually it plays a similar role as the r.h.s. second term
in the vortical transition density (\ref{vtd}). Both prescriptions, (\ref{vtd}) and (\ref{12}),
have the same intention but, being applied to to different quantities
(transition densities and operators), lead to formally different recipes.
While (\ref{vtd})  excludes the $j_{\lambda \:\lambda-1}(r)$ terms,
the recipe (\ref{12}) gives an exact compensation of the
lowest-order $k$-terms in the final vortical operator, see
the derivation below. Both (\ref{vtd}) and (\ref{12}) remind the r.h.s. of
the HD expression (\ref{vor_v_j}). They
are compared and discussed in more detail in Appendix \ref{Sec:_constraint}.

By using (\ref{12}), the vortical operator is defined as
\begin{equation}\label{vort_oper_E-S}
\hat{M}_{\text vor}(E\lambda\mu, k) = \hat{M}(E\lambda\mu, k) - \hat{M}_S(E\lambda\mu, k)
\end{equation}
i.e. as a  difference of the electric operator (\ref{ME_oper2})
and the subsidiary operator
\begin{eqnarray}\label{S_oper1}
\hat{M}_S(E\lambda\mu, k) &=&
i \frac{(2\lambda + 1)!!}
{k^{\lambda}\sqrt{\lambda(\lambda+1)}}
\\ \nonumber
&& \cdot\int d^3r \: [\: j_{\lambda}(kr)\:
\vec{Y}_{\lambda \lambda \mu}(\hat{\vec r}) \:]
\cdot
\: [\vec{\nabla} \hat{\rho}(\vec{r})  \times \vec{r}] \; .
\end{eqnarray}
The latter may be also written in the forms
\begin{eqnarray}
\label{S_oper2}
&&\hat{M}_S(E\lambda\mu, k) =
\nonumber
\\
&=&- \frac{(2\lambda + 1)!!}
{k^{\lambda}\sqrt{\lambda(\lambda+1)}}
\int d^3r \:  \hat{\rho}(\vec{r}) \: j_{\lambda}(kr)\:
\hat{\vec{l}} \cdot \vec{Y}_{\lambda \lambda \mu}(\hat{\vec r})
\\
\label{S_oper3}
&=&
- \frac{(2\lambda+1)!!}{k^{\lambda}}
\int d^3r \hat{\rho}(\vec{r}) j_{\lambda}(kr)
Y_{\lambda \mu}(\hat{\vec r})
\\
\label{S_oper4}
&=&
- i\frac{(2\lambda + 1)!!}{ck^{\lambda+1}}
\int d^3r j_{\lambda}(kr) Y_{\lambda \mu}(\hat{\vec r})
 [ \vec{\nabla} \cdot \hat{\vec{j}}_{\text nuc}(\vec{r})]
\end{eqnarray}
The form (\ref{S_oper4}) is obtained by using the operator continuity
equation
\begin{equation}\label{j_rho}
\vec{\nabla} \cdot \hat{\vec{j}}_{\text nuc} = -\frac{i}{\hbar} [\hat{H},
\hat{\rho}] = -ikc\hat{\rho} \; .
\end{equation}

In the long-wavelength approximation ($k\rightarrow 0$), we
keep only the first and second terms in the expansion of the spherical
Bessel function
\begin{equation}
j_{\lambda}(kr) = \frac{(kr)^{\lambda}}{(2 \lambda +1)!!} \:
[\:1 - \frac{(kr)^2}{2(2 \lambda +3)} + \ldots \:]
\label{18}
\end{equation}
and thus get for the electric and subsidiary operators
\begin{eqnarray}\label{E+ktor}
\hat{M}(E\lambda\mu, k) &\approx& \hat{M}(E\lambda\mu)
+ k\:\hat{M}_{\text tor}(E\lambda\mu) \;,
\\
\label{E-kcom}
\hat{M}_S (E\lambda\mu, k) &\approx& \hat{M}(E\lambda \mu)
- k \hat{M}_{\text com}(E\lambda\mu) \; ,
\end{eqnarray}
where
\begin{eqnarray}\label{E_oper_main}
\hat{M}(E\lambda \mu) &=&
\frac{i}{kc} \int d^3r \: \hat{\vec{j}}_{\text nuc}(\vec{r})
\cdot \vec{\nabla} (r^{\lambda} Y_{\lambda \mu}(\hat{\vec r}))
\\ \nonumber
&=& -\frac{i}{kc} \int d^3r \: (\vec{\nabla}
\cdot \hat{\vec{j}}_{\text nuc}(\vec{r}))
 r^{\lambda} Y_{\lambda \mu}(\hat{\vec r})
\\ \nonumber
&=& - \int d^{3}r \:\hat{\rho}(\vec{r})\: r^{\lambda} Y_{\lambda
\mu}(\hat{\vec r})
\end{eqnarray}
is the familiar electrical operator in the long-wavelength limit
(lowest order term) and
\begin{eqnarray}\label{tor_rel_1}
\hat{M}_{\text tor} (E\lambda \mu)
&=& \frac{i}{2c(\lambda+1)(2\lambda +3)}
\\
&&\cdot\int d^3r \hat{\vec{j}}_{\text nuc}(\vec{r}) \cdot [ \:\vec{\nabla}
\times \:(\vec{r} \times \vec{\nabla}) r^{\lambda +2} Y_{\lambda \mu}(\hat{\vec
r}) \:] \nonumber
\\
\label{tor_rel_2}
&=& \frac{i}{c(\lambda +1)}
\int d^3r \hat{\vec{j}}_{\text nuc}(\vec{r})
 \cdot {\vec r} r^{\lambda} Y_{\lambda\mu}(\hat{\vec r})
\\ \nonumber
&+& k \frac{\lambda+3}{2(\lambda +1) (2\lambda +3)}
\int d^3r \hat{\rho}(\vec r) r^{\lambda +2}Y_{\lambda\mu}(\hat{\vec r}) \; ,
\end{eqnarray}
\begin{eqnarray}\label{com__rel_1}
\hat{M}_{\text com} (E\lambda \mu)
&=& \frac{i}{2c (2\lambda +3)}
\\
&& \cdot\int d^3r \hat{\vec{j}}_{\text nuc}(\vec{r}) \cdot \:\vec{\nabla}  [\:
r^{\lambda +2} Y_{\lambda \mu}(\hat{\vec r}) \:] \nonumber
\end{eqnarray}
are toroidal and compressional operators, respectively.
Note that both toroidal expressions (\ref{tor_rel_1}) and
(\ref{tor_rel_2}) involve the function $r^{\lambda+2}Y_{\lambda\mu}(\hat{\vec
r})$ thus manifesting the relation between TM and CM. In (\ref{tor_rel_2}), the
second term precisely gives the CM operator \cite{Kv03}.

In Eq. (\ref{vort_oper_E-S}), the lowest-order $k$-terms from
(\ref{E+ktor}) and (\ref{E-kcom}) exactly compensate each other
and so we get
\begin{equation}\label{vor_oper_k}
\hat{M}_{\text vor} (E\lambda\mu, k) =  k \:[ \hat{M}_{\text tor}(E\lambda \mu)
+ \hat{M}_{\text com}(E\lambda\mu)] \;.
\end{equation}
Using the definition,
$\hat{M}_{\text vor} (E\lambda\mu, k) = k \hat{M}_{\text vor} (E\lambda\mu)$,
we finally come to the relation
\begin{equation}\label{vor=tor+com}
\hat{M}_{\text vor} (E\lambda\mu) = \hat{M}_{\text tor}(E\lambda\mu)
+ \hat{M}_{\text com}(E\lambda\mu) \; ,
\end{equation}
where
\begin{widetext}
\begin{eqnarray}
 \hat M_{\text vor}(E\lambda\mu)
  &=&
  -\frac{i}{c(2\lambda+3)}\sqrt{\frac{2\lambda+1}{\lambda+1}}
 \cdot \int d^3r
 \hat{\vec{j}}_{\text nuc}(\vec r)
 r^{\lambda+1}
 \vec Y_{\lambda \lambda+1 \mu} (\hat{\vec r}) \; ,
\label{vort_oper_k0}
\\
 {\hat M}_{\text tor}(E\lambda\mu)
 &=&
 -\frac{i}{2c}
 \sqrt{\frac{\lambda}{2\lambda+1}}
 \int d^3r \:
 \hat{\vec j}_{\text nuc}(\vec r)
 \cdot r^{\lambda+1}
 \left[ \vec Y_{\lambda\lambda-1\mu}(\hat{\vec r}) +
 \sqrt{\frac{\lambda}{\lambda+1}}\frac{2}{2\lambda+3}
  \vec Y_{\lambda\lambda+1\mu}(\hat{\vec r})
 \right]
\label{tor_oper}
\\
  &=&
 -\frac{1}{2c}
 \sqrt{\frac{\lambda}{\lambda+1}}\,\frac{1}{2\lambda+3}
 \int d^3r \: r^{\lambda+2}\vec{Y}_{\lambda\lambda\mu}(\hat{\vec r})
 \cdot
  \left(\vec{\nabla} \times \hat{\vec j}_{\text nuc}(\vec r)\right)
  \; ,
\nonumber
\\
  \hat{M}_{\text com}(E\lambda\mu)
  &=&
  \frac{i}{2c}\:  \sqrt{\frac{\lambda}{2\lambda+1}}
\int d^3r \: \hat{\vec{j}}_{nuc}(\vec{r}) \cdot  r^{\lambda+1} \left[
\vec{Y}_{\lambda \lambda-1 \mu}(\hat{\vec r}) -
\sqrt{\frac{\lambda+1}{\lambda}} \: \frac{2}{2\lambda+3} \vec{Y}_{\lambda
\lambda+1 \mu}(\hat{\vec r}) \right] \label{comp_op_CD}
\\
  &=&
 - \frac{i}{2c}\:\frac{1}{2\lambda+3}
 \int d^3r \: r^{\lambda+2}{Y}_{\lambda\mu}(\hat{\vec r})
  \left(\vec{\nabla} \cdot \hat{\vec j}_{\text nuc}(\vec r) \right) \; .
\nonumber
\end{eqnarray}
\end{widetext}
Here, the TM and VM operators
are the same as in (\ref{tor_rel_1})-(\ref{com__rel_1}) but are written
in the forms convenient for the comparison with the VM operator.
Besides these forms demonstrate the
$\vec{\nabla} \times \hat{\vec j}_{\text nuc}(\vec r)$ and
$\vec{\nabla} \cdot \hat{\vec j}_{\text nuc}(\vec r)$ origin of the
TM and CM operators, respectively.

The expression  for the VM operator (\ref{vort_oper_k0}) and the relation
(\ref{vor=tor+com}) between VM, TM, and CM operators represent
the main formal results of the present paper.

Following (\ref{vor=tor+com}), the operators $\hat{M}_{\text vor} (E\lambda\mu)$,
$\hat{M}_{\text tor}(E\lambda\mu)$, and  $\hat{M}_{\text com}(E\lambda\mu)$
are of the same second order by $k$. They are given in
(\ref{vort_oper_k0})-(\ref{comp_op_CD}) in the
current-dependent form. Using the continuity equation (\ref{j_rho}),
the current-dependent CM operator (\ref{comp_op_CD}) is straightforwardly
transformed to the familiar density-dependent CM operator \cite{Semenko_81}
\begin{equation}\label{comp_op_DD}
\hat{M}'_{\text com}(E\lambda\mu)=  \frac{1}{2(2\lambda+3)}\int d^3r
\hat{\rho}(\vec r) r^{\lambda+2} Y_{\lambda\mu}(\hat{\vec r})
\end{equation}
as
\begin{equation}\label{oper_com_com'}
\hat{M}_{\text com}(E\lambda\mu)= -k \hat{M}'_{\text com}(E\lambda\mu) .
\end{equation}

Note that the relation (\ref{vor=tor+com}) requires
the compensation of the terms
$\sim\vec{Y}_{\lambda\lambda-1 \mu}(\hat{\vec r})$ in the TM and CM
operators. Thus a simultaneous use of these two operators is
obligatory.  The VM operator includes only
$\vec{Y}_{\lambda\lambda+1\mu}(\hat{\vec r})$ and so its matrix
elements are determined by the current transition density
$j_{\lambda\:\lambda+1}(r)$, as requested
in \cite{Ra87}. It is easy to check that
$\hat{M}_{\text vor}(E\lambda\mu)$ gives the transition vorticity
(\ref{ctdm}) and so reproduces the results \cite{Ra87}.

The above formalism was derived for the case when the system
is excited by the external electric field, i.e. for the excitation energy
$\omega=\hbar ck > 0$. The case of de-excitation is easily obtained by
replacement $k \to -k$ in the continuity equations (\ref{ce_ctd}) and
(\ref{j_rho}), k-dependent terms in (\ref{12}),
(\ref{tor_rel_2}), (\ref{oper_com_com'}), and equations of Appendix C. The sign
of $\hat{M}_S(E\lambda\mu, k)$ and density-dependent $\hat{M}(E\lambda\mu)$ is
changed as well.

\subsection{Dipole VM, TM, and CM  operators}

The VM, TM, and CM are usually studied for the electric $I^{\pi}=1^-$ states
\cite{Pa07}. Then the operators (\ref{vort_oper_k0})-(\ref{comp_op_CD}) are
reduced to
\begin{eqnarray}
\label{vort_dip_oper} && \hat M_{\text vor}(E1\mu) = -\frac{i}{5c}
\sqrt{\frac{3}{2}} \int d^3r  \hat{\vec{j}}_{\text nuc}(\vec r) r^{2} \vec Y_{1
2 \mu}(\hat{\vec r}) ,
\\
\label{tor_dip_oper} && \hat M_{\text tor}(E1\mu) = -\frac{i}{2\sqrt{3}c} \int
d^3r \hat{\vec j}_{\text nuc}(\vec r)
\\ \nonumber
&&\quad  \cdot \; [\frac{\sqrt{2}}{5}r^2\vec Y_{12\mu}(\hat{\vec r}) + (r^2
-\delta_{T,0}\langle r^2\rangle_0) \vec Y_{10\mu}(\hat{\vec r})] ,
\end{eqnarray}
\begin{eqnarray}
\label{comp_dip_op}
&& \hat M_{\text com}(E1\mu) = -\frac{i}{2\sqrt{3}c} \int
d^3r \hat{\vec j}_{\text nuc}(\vec r)
\\ \nonumber
&&\quad  \cdot \; [\frac{2\sqrt{2}}{5} r^2\vec Y_{12\mu}(\hat{\vec r}) - (r^2 -
\delta_{T,0}\langle r^2\rangle_0) \vec Y_{10\mu}(\hat{\vec r})] ,
\\
\label{comp_oper} && \hat M'_{\text com}(E1\mu) =  \frac{1}{10} \int d^3r
\hat{\rho}(\vec r)
\\ \nonumber
&&\quad \cdot \; [r^3-\delta_{T,0}\frac{5}{3}\langle r^2\rangle_0 r]
Y_{1\mu}(\hat{\vec r}) ,
\end{eqnarray}
where $\langle r^2 \rangle_0^{\mbox{}}=\int d^3r \rho_0(\vec{r}) r^2/A$ is the
ground-state squared radius.

In (\ref{tor_dip_oper})-(\ref{comp_oper}), the terms $\sim \vec
Y_{10\mu}(\hat{\vec r}), \; Y_{1\mu}(\hat{\vec r})$ include the center of mass
corrections (c.m.c.) for T=0 excitations \cite{Semenko_81}. In the TM and
current-dependent CM operators, the c.m.c. have the same magnitude. For the VM,
the c.m.c. is zero, see discussion in Appendix \ref{Sec:_spurious}.

The expression in the square brackets of the TM operator (\ref{tor_dip_oper})
can be written as \cite{Semenko_81}
\begin{equation}\label{tor_vel_E1}
\vec{\nabla} \times (\vec r \times \vec{\nabla} ) (r^3-\frac{5}{3}\langle
r^2\rangle_0^{\mbox{}} \; r)Y_{1\mu}(\hat{\vec r}) \; ,
\end{equation}
which justifies a close relation between TM and CM.

\subsection{Discussion of  VM, TM, and CM operators and vorticity criteria}

As shown above, the CM operator may be presented in the current-dependent
(\ref{comp_op_CD}) and density-dependent (\ref{comp_op_DD}) forms.  To
the best of our knowledge, the former has not yet been used in the
literature.  Since the spin current $j_m$ is a curl of the
magnetization, it does not contribute to the continuity equation and
CM operator.
Though the current-dependent form (\ref{comp_op_CD}) of CM formally involves
$j_m$, its contribution is annihilated
by $\vec{\nabla} \cdot \hat{\vec j}_{\text nuc}(\vec r)$
or the corresponding combinations of vector spherical harmonics.
So actually both forms, (\ref{comp_op_CD}) and
(\ref{comp_op_DD}), of the CM operator do not depend on $j_m$.

Both CM operators, (\ref{comp_op_CD}) and (\ref{comp_op_DD}),
are obtained through $\vec{\nabla} \cdot \hat{\vec{j}}_{\text nuc}(\vec r)$,
which suggests their vorticity-free character.
This is confirmed by the form of the CM velocity which, following the
prescription \cite{Suzuki_Rowe_NPA}, reads
\begin{equation}\label{v_CM}
  \vec{v}_{\text com}(\vec{r})
  \propto
  \vec{\nabla}\, r^{\lambda +2} Y_{\lambda\mu}(\hat{\vec r})
\end{equation}
and so gives $\vec{\nabla} \times  \vec{v}_{\text com}(\vec{r})$=0.

The current-dependent CM operator (\ref{comp_op_CD})
includes the $\vec{Y}_{\lambda \lambda+1 \mu}(\hat{\vec r})$
contribution which might be considered as an indicator
of a vortical part. Indeed
$\vec{Y}_{\lambda \lambda+1 \mu}(\hat{\vec r})$
leads to the current component
$j_{\lambda \: \lambda+1}(r)$ which, following \cite{Ra87}, is
responsible for the  vorticity. However, for the CM this is misleading.
The velocity (\ref{v_CM}) can be straightforwardly cast into the form
\begin{eqnarray}\label{v_CM_Y}
  \vec{v}_{\text com}(\vec{r}) && \propto
  \sqrt{\frac{\lambda}{2\lambda+1}} r^{\lambda+1}
  \\
&& \cdot\left[ \vec{Y}_{\lambda \lambda-1 \mu}(\hat{\vec r}) -
\sqrt{\frac{\lambda+1}{\lambda}} \: \frac{2}{2\lambda+3} \vec{Y}_{\lambda
\lambda+1 \mu}(\hat{\vec r}) \right] \nonumber
\end{eqnarray}
whose curl is zero despite the
$\vec{Y}_{\lambda \lambda+1 \mu}(\hat{\vec r})$ term.
So, following the HD criterion (\ref{vort}), the
appearance of a term
$\vec{Y}_{\lambda \lambda+1 \mu}(\hat{\vec r})$ in the mode operator
and current is not yet a definitive signature of the vorticity.

Altogether, we see apparent differences between two possible criteria for
the vorticity: i) the HD condition (\ref{vort}) in terms of
velocities, $\vec{\nabla} \times \vec{v} (\vec{r}) \ne 0,$ and ii) the
condition \cite{Ra87} in terms of transition current density components,
$j_{\lambda \: \lambda+1}(r) \ne 0$. As shown above for CM example, a mode
which is fully vorticity-free in the HD definition can have a
substantial vorticity of the sort \cite{Ra87}. The difference between these
criteria might be understood if we take into account that the vorticity density
$w_{\lambda \lambda}(r) \propto \:j_{\lambda \: \lambda+1}(r)$ was derived in
\cite{Ra87}, first of all, as a quantity completely unconstrained by the charge
conservation rather than a purely vortical value in the HD sense.

In fluid-dynamical and HD models \cite{Bast93,Mi06},
the velocity fields are chosen in form of the relevant external fields
exciting the proper modes. Following this practice, the
TM  velocity reads
\begin{eqnarray}\label{v_tor}
 \vec{v}_{\text tor}(\vec{r})&& \propto
\vec{\nabla} \times (\vec r \times \vec{\nabla} ) r^{\lambda +2}
Y_{\lambda\mu}(\hat{\vec r})
\\
\label{v_tor1}
&&  =
 i\sqrt{\frac{\lambda}{2\lambda +1}} (\lambda +1)(2\lambda +3)r^{\lambda+1}
 \\
&&  \cdot\left[ \vec{Y}_{\lambda \lambda-1 \mu}(\hat{\vec r}) +
\sqrt{\frac{\lambda}{\lambda +1}} \: \frac{2}{2\lambda+3} \vec{Y}_{\lambda
\lambda+1 \mu}(\hat{\vec r}) \right] \; . \nonumber
\end{eqnarray}
It is easy to check that $\vec{\nabla} \times \vec{v}_{\text tor} \propto
r^{\lambda } \vec{Y}_{\lambda \lambda \mu}(\hat{\vec r})$ and so TM
carries the HD vorticity.

The VM velocity constructed in the same manner is
\begin{equation}
\label{v_vor}
 \vec{v}_{\text vor}(\vec{r}) \propto r^{\lambda+1}
\vec{Y}_{\lambda \lambda+1 \mu}(\hat{\vec r}) \; .
\end{equation}
It has the similar nonzero curl $\vec{\nabla} \times \vec{v}_{\text vor} \propto
r^{\lambda } \vec{Y}_{\lambda \lambda \mu}(\hat{\vec r})$. So the VM has
the vorticity of both HD and Ref. \cite{Ra87} sorts.

\section{Calculation scheme}
\label{sec:calc_scheme}

The calculations of the excitation modes were performed within the separable
random-phase approximation (SRPA) model using the Skyrme energy functional
\cite{nest_PRC_02,nest_PRC_06}. SRPA was earlier successfully applied
to description of electric
\cite{nest_PRC_02,nest_PRC_06,nest_IJMPE_07,nest_IJMPE_08,nest_PRC_08}
and magnetic \cite{Ve09,Nest_JPG_10,Nest_IJMPE_10} GR in spherical and
deformed nuclei.  The approach was also used for the exploration of E1
strength near particle emission thresholds \cite{Kvasil_IJMPE_09}.
SRPA is {\it fully self-consistent} in the sense that both the static
mean-field and factorized residual interaction are derived from the same
Skyrme energy functional \cite{Skyrme,Vau,Engel_75}. The functional,
$\mathcal{E}_{\mathrm{Sk}} (\rho, \tau, \vec{J}, \vec{j}, \vec{s},
\vec{T})$, includes time-even (nucleon $\rho$, kinetic-energy $\tau$,
spin-orbit $\vec{J}$) and time-odd (current $\vec{j}$, spin $\vec{s}$,
vector kinetic-energy $\vec{T}$) densities. It also involves pairing
(surface and volume), Coulomb (direct and exchange), and
c.m.c. terms \cite{nest_PRC_02,nest_PRC_06,Ve09}. The
Galilean invariance of the functional is maintained in SRPA.
The tensor spin-orbit contribution is involved through the squared
spin-orbit densities $\vec{J}^2$.
All the functional terms are kept in the mean field
and residual interaction.

The SRPA expands the RPA residual interaction self-consistently into a
sum of separable terms, which dramatically reduces the computational effort
while keeping the accuracy of the full (non-separable)
RPA \cite{nest_PRC_02,nest_PRC_06}. This makes SRPA extremely useful
for systematic calculations and tasks with very large
configuration space, e.g.  for description of GR in heavy deformed
nuclei.

For GR studies, the computational expense can be even more reduced by
a direct evaluation of the strength function, thus avoiding the
solution of RPA equations for a large multitude of individual
states. The SRPA strength function works with the Lorentz weight and has
very simple form \cite{nest_PRC_02,nest_PRC_06}. In the present study,
the strength function for electric dipole modes reads
\begin{equation}\label{eq:strength_function}
  S_{\alpha}(E1; \omega) = (2\pi)^2
   \sum_{\mu = 0,\pm 1}
   \sum_{\nu}
  |\langle\Psi_\nu|\hat{M}_{\alpha}(E1\mu)|\Psi_0\rangle|^2
  \zeta(\omega - \omega_{\nu})
\end{equation}
where
\begin{equation}
  \zeta(\omega - \omega_{\nu}) =
  \frac{1}{2\pi}\frac{\Delta}{(\omega- \omega_{\nu})^2+\frac{\Delta^2}{4}}
\end{equation}
is the Lorentz weight with the smoothing width $\Delta$ and
$\hat{M}_{\alpha}(E1\mu)$ is the electric dipole transition operator
whose type is determined by the index $\alpha = \{\text{vor, tor, com,
com'}\}$. Further, $\Psi_0$ is the ground state, $\nu$ runs over
the RPA spectrum with eigen-frequencies $\omega_{\nu}$ and
eigen-states $|\Psi_\nu\rangle$.  The Lorentz smoothing uses
a width $\Delta$=1 MeV to simulate broadening effects beyond RPA
(escape widths and coupling to complex configurations). The
explicit form of the strength function (\ref{eq:strength_function}),
which does not directly involve
the RPA solutions, is given elsewhere \cite{nest_PRC_02,nest_PRC_06}.

In general, the VM, TM, and CM appear at many different
multipolarities $\lambda\mu$. Here we explore them in the
$I^{\pi}=1^-$ excitations of the doubly-magic spherical nucleus
$^{208}$Pb. In this nucleus the pairing is absent. The calculations
use the Skyrme parameterization SLy6 \cite{Sly6}, which provides a
satisfactory description of the E1(T=1) GR in heavy nuclei
\cite{nest_PRC_08}. For this parameterization, the tensor
spin-orbit contribution is omitted.

The calculations employ a
cylindrical coordinate-space grid with the mesh size 0.7 fm. A large
single-particle basis is used. The particle-hole $1^-$ pairs
extend up to $\sim 65$ MeV and, for E1(T=1) excitations,
the energy-weighted sum rule with the isovector effective mass
is exhausted by $\sim 95\%$
\cite{nest_PRC_08}.

The calculations involve both convection $j_c$ and magnetization (spin)
$j_m$ parts of the nuclear current, see Appendix \ref{Sec:_density_oper}
for more details. For $j_m$, the isoscalar
$g^{n,p}_s(T=0)=(g^{p}_s + g^{n}_s)/2=0.88\varsigma$ and
isovector $g^{n,p}_s(T=1)=(g^{n}_s - g^{p}_s)/2=-4.70\varsigma$
gyromagnetic factors are used, where
$g^{p}_s = 5.58 \varsigma$ and $g^{n}_s = - 3.82 \varsigma$
are bare proton and neutron $g$-factors and $\varsigma$=0.7
is a quenching parameter taking approximately into account the
meson degrees of freedom \cite{Harakeh_book_01}.  It is easy to see that
$|g^{n,p}_s(T=1)|\gg |g^{n,p}_s(T=0)|$ and so the main effect of
the spin nuclear current is expected for T=1 modes.

The proton and neutron ($q = n, p$) RPA velocity fields are determined in cylindrical
coordinates $(z,\rho)$ as
\begin{equation}\label{vel}
{\vec v}^q_{\nu}({z,\rho})= \frac{\delta \vec{j}_{\nu q}({z,\rho})}{\rho^q_0 (z,\rho)}
\end{equation}
where
\begin{equation}\label{current_tr_dens}
\delta{\vec j}_{\nu q}(z,\rho)= \sum_{ij \in q}
\langle ij|\hat{\vec j}_{nuc}^{q}|0\rangle (c^{\nu q-}_{ij}-c^{\nu q+}_{ij})
\end{equation}
is the current transition density for the RPA $\nu$-state with the normalized
forward and backward particle-hole (1ph) amplitudes $c^{\nu q-}_{ij}$ and
$c^{\nu q+}_{ij}$. Further, $\rho^q_0 (z,\rho)$ is the proton/neutron
ground state density.

The spurious c.m. admixtures are avoided by using the prescriptions
from the Appendix \ref{Sec:_spurious}. The SRPA equations and some important points,
e.g. a choice of the generating operators for the separable
expansion, are sketched in the Appendix \ref{Sec:_SRPA}.

\section{Results and discussion}
\label{sec:results}

\begin{figure}[t]
\includegraphics[width=7.5cm]{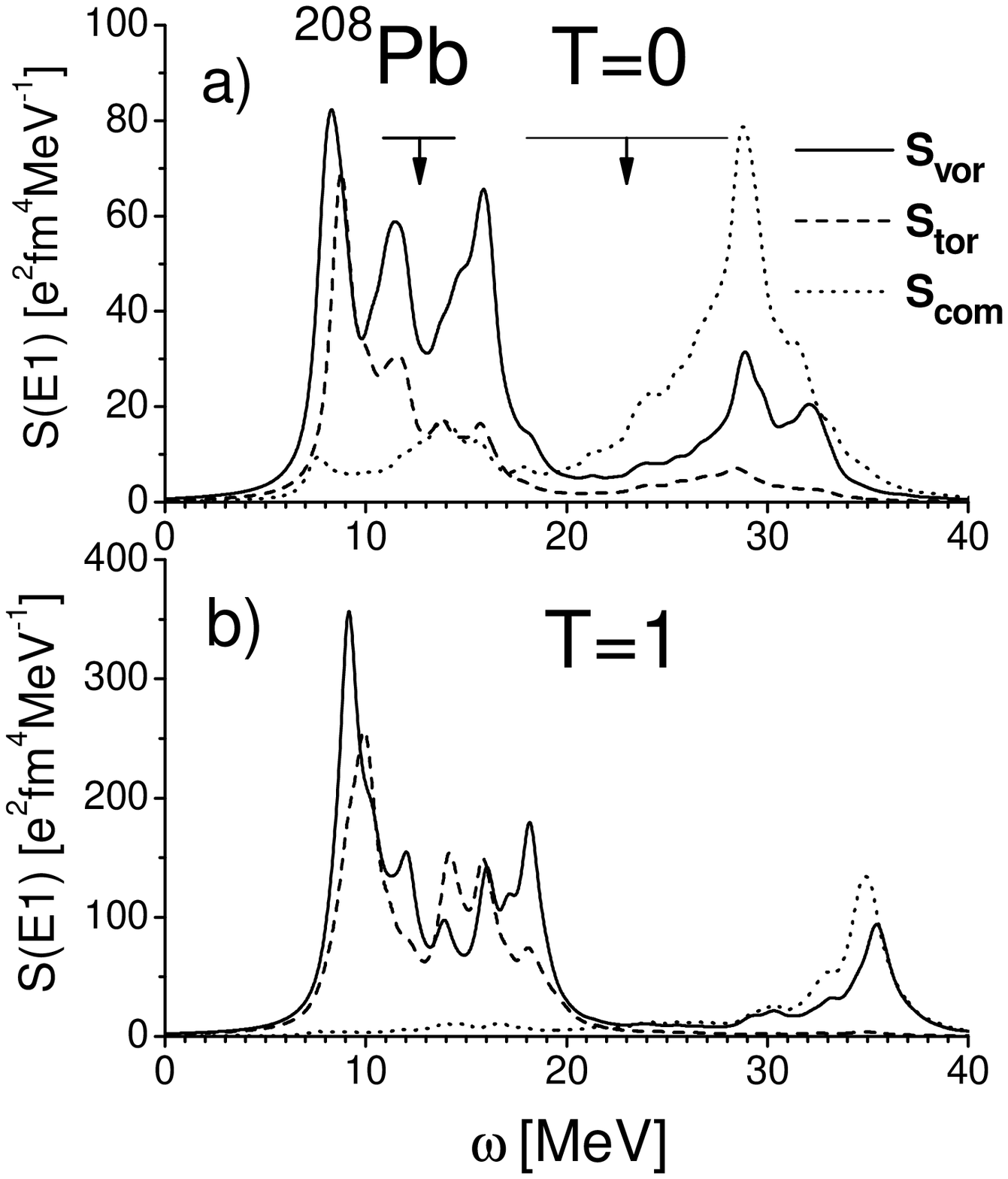}
\caption{ \label{fig:fig1} Isoscalar (T=0) and isovector (T=1) vortical,
toroidal, and compression dipole modes in $^{208}$Pb, calculated with the SLy6
parameterization. The total nuclear current $j_{nuc}$ is used. The CM is
computed with the operator $\hat M_{com}$ from (\ref{comp_dip_op}).  The lines
with the arrows indicate widths  and energy centroids   of the low-energy and
high-energy branches of isoscalar E1 excitations observed in $(\alpha,
\alpha')$ reaction \protect\cite{Uchida_PLB_03,Uchida_PRC_04}.}
\end{figure}

\begin{figure}[t]
\includegraphics[width=7.1cm]{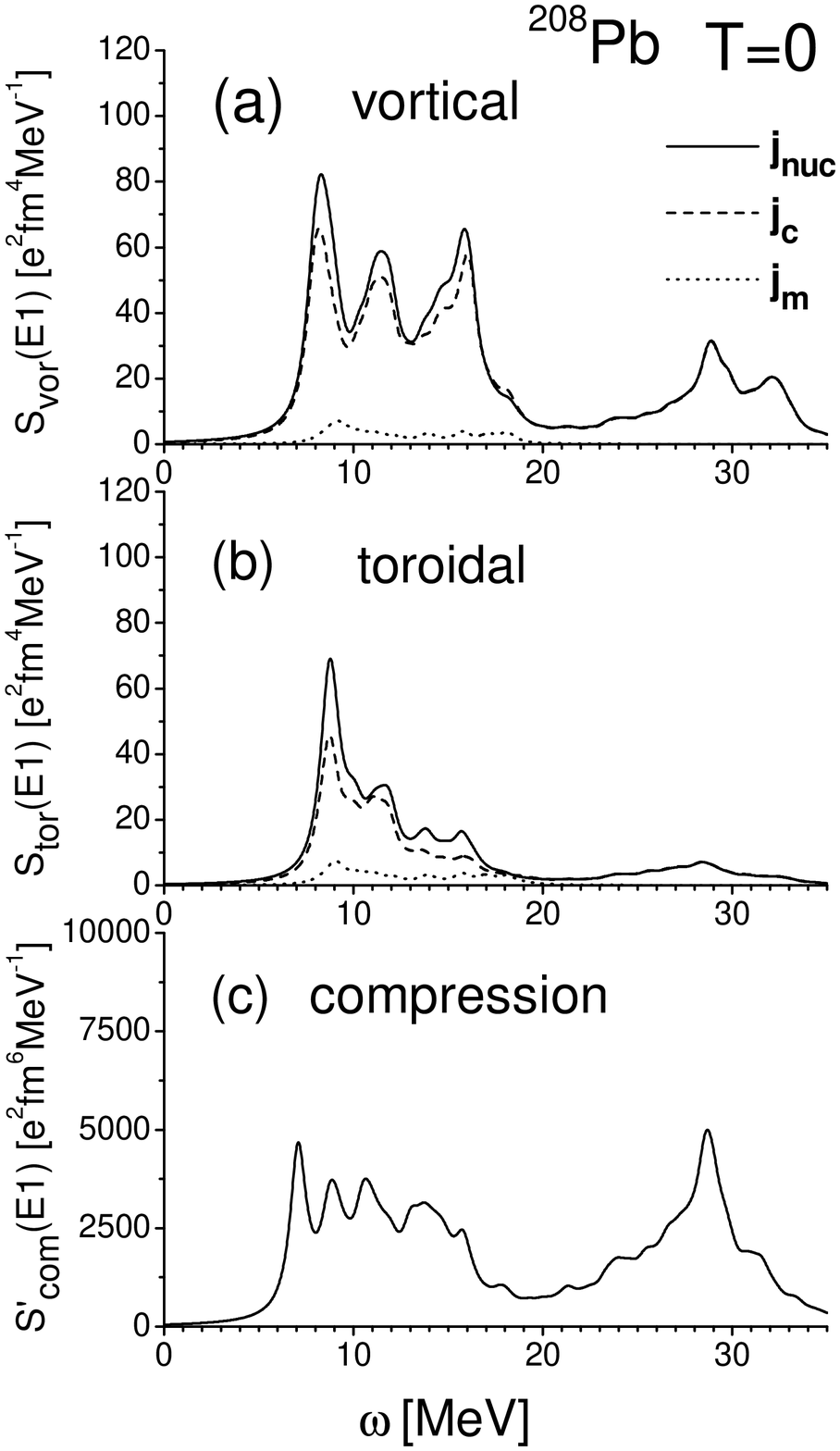}
\vspace{0.2cm} \caption{ \label{fig:fig2} Calculated isoscalar (T=0) vortical,
toroidal, and compression dipole modes in $^{208}$Pb. For the VM and TM, the
strengths with the total $j_{nuc}$, convection $j_c$, and magnetization $j_m$
current contributions to the transition operators are shown.}
\end{figure}

\begin{figure}[t]
\includegraphics[width=7.1cm]{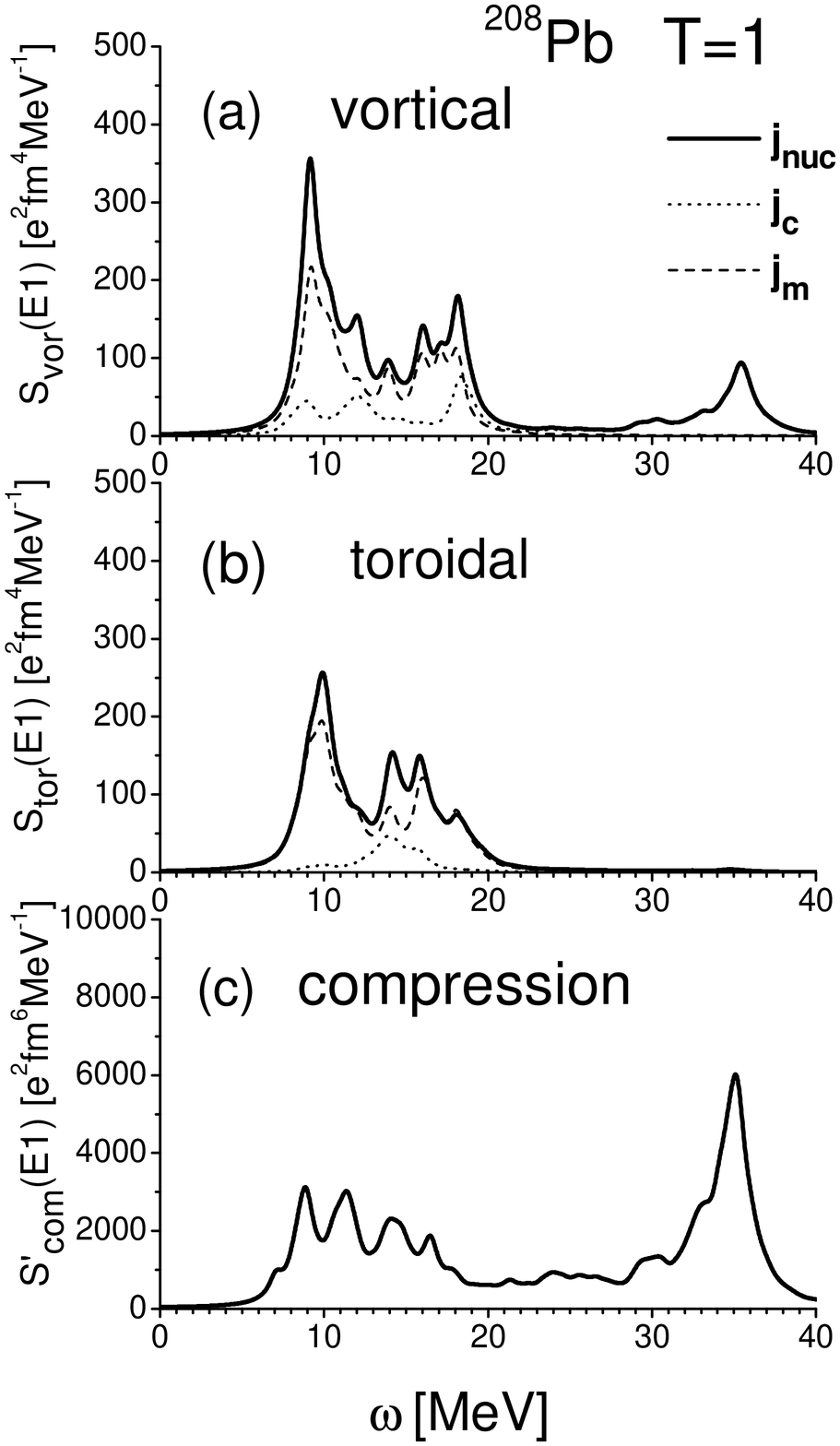}
\caption{ \label{fig:fig3} The same as in Fig. \protect\ref{fig:fig2} but for
the T=1 modes.}
\end{figure}

\begin{figure}[t]
\includegraphics[width=7.3cm]{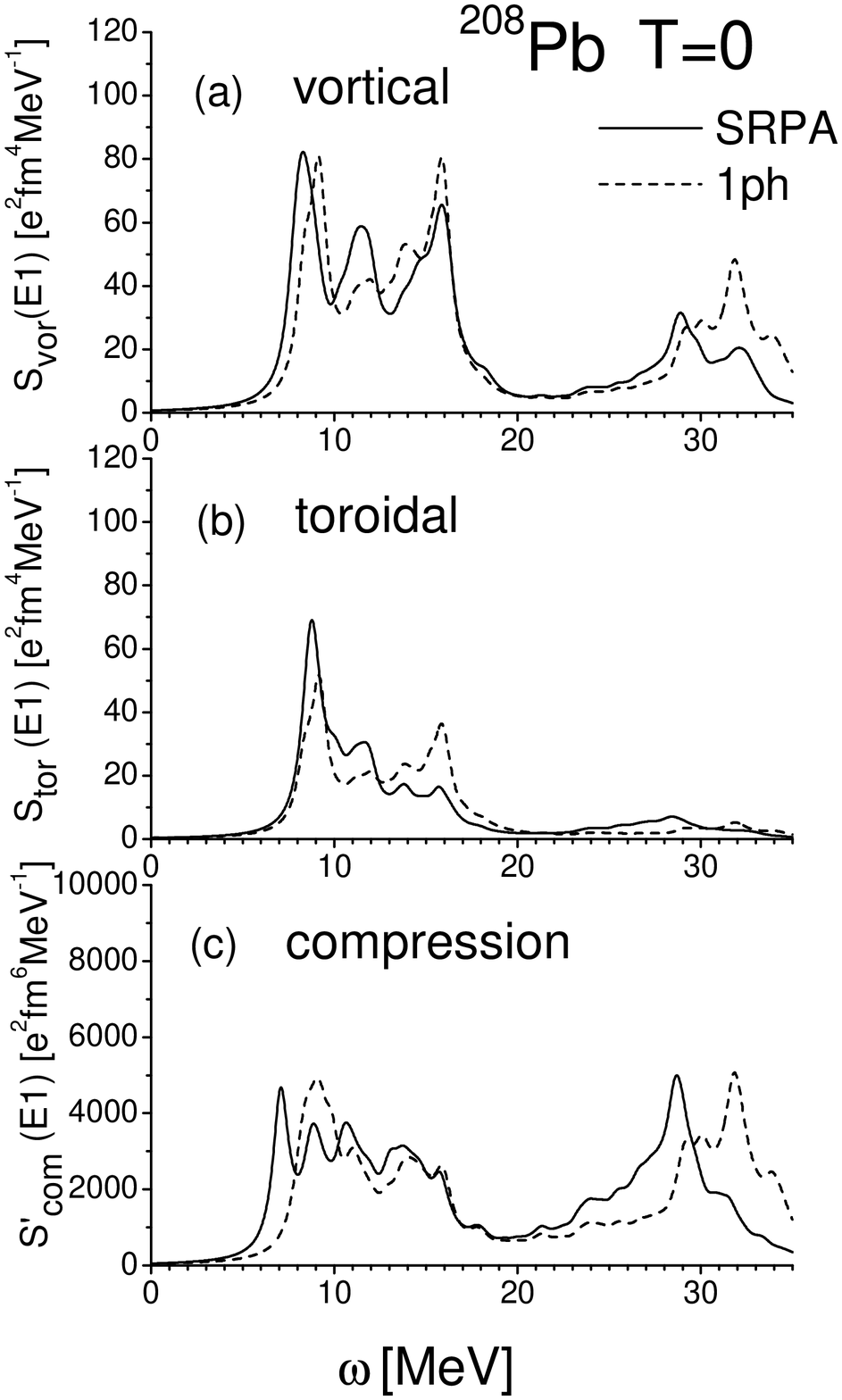}
\caption{
\label{fig:fig4}
Comparison of SRPA and 1ph strengths for the T=0 modes.
For VM and TM, the strengths are computed with the total nuclear current
$j_{nuc}$.}
\end{figure}

\begin{figure}[t]
\includegraphics[width=7.2cm]{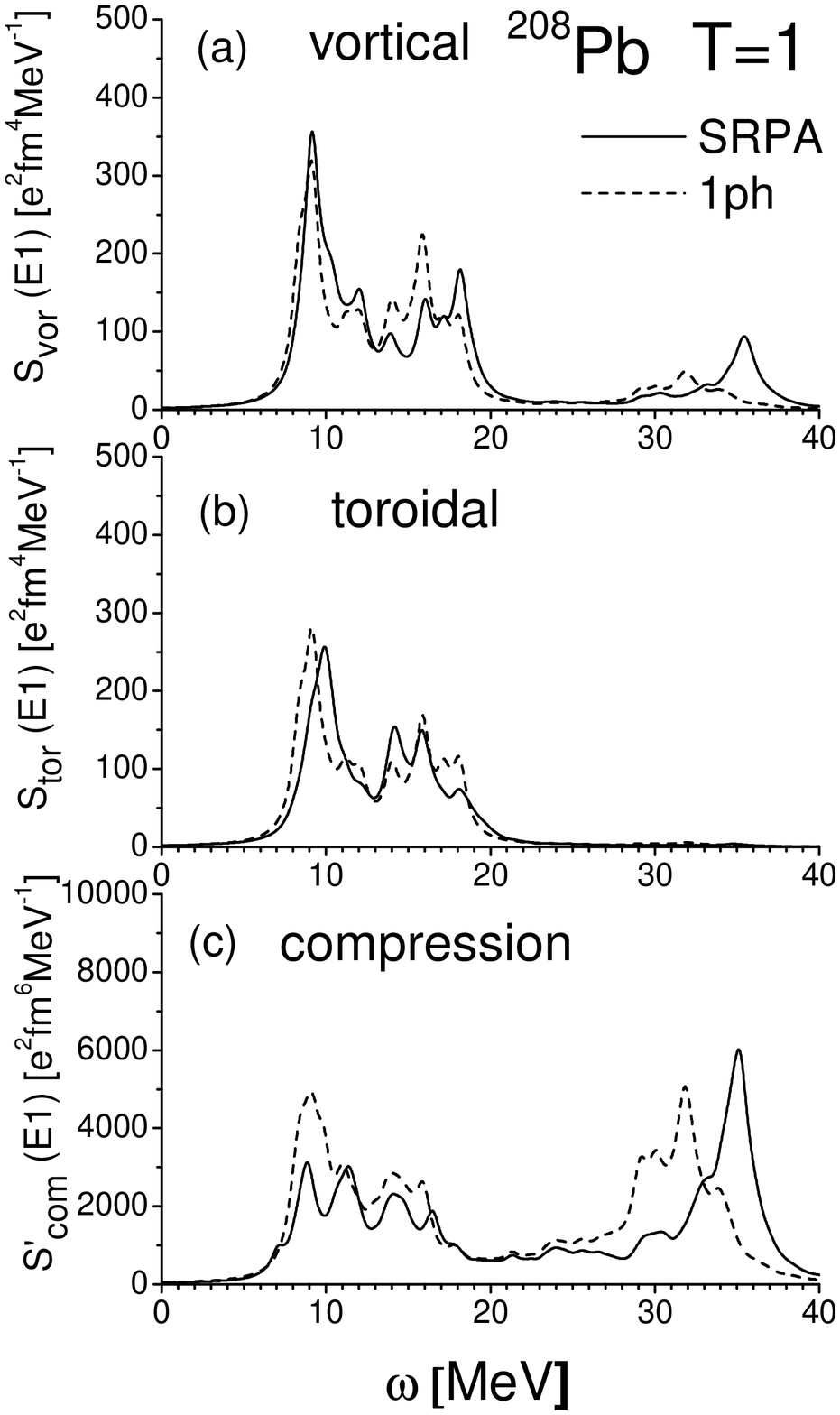}
\caption{
\label{fig:fig5}
The same as in Fig. \ref{fig:fig4} but for the T=1 modes.}
\end{figure}

\begin{figure}
\includegraphics[width=7.3cm]{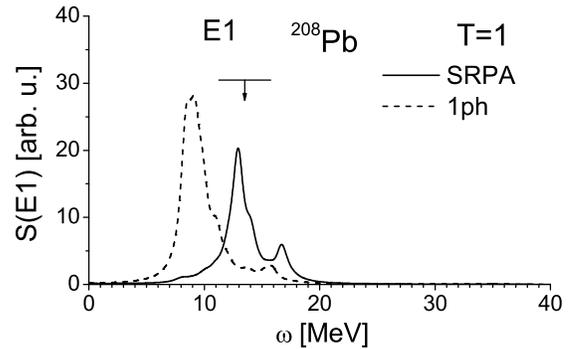}
\caption{\label{fig:fig6}
SRPA and 1ph strengths for the E1(T=1) GDR. Like for other modes,
the strength functions are plotted without the energy weight.
The experimental width and energy
\protect\cite{varlamov_atlas_99} are shown by the horizontal line and arrow,
respectively.}
\end{figure}

\begin{figure}[t]
\includegraphics[height=7.5cm,width=7.5cm]{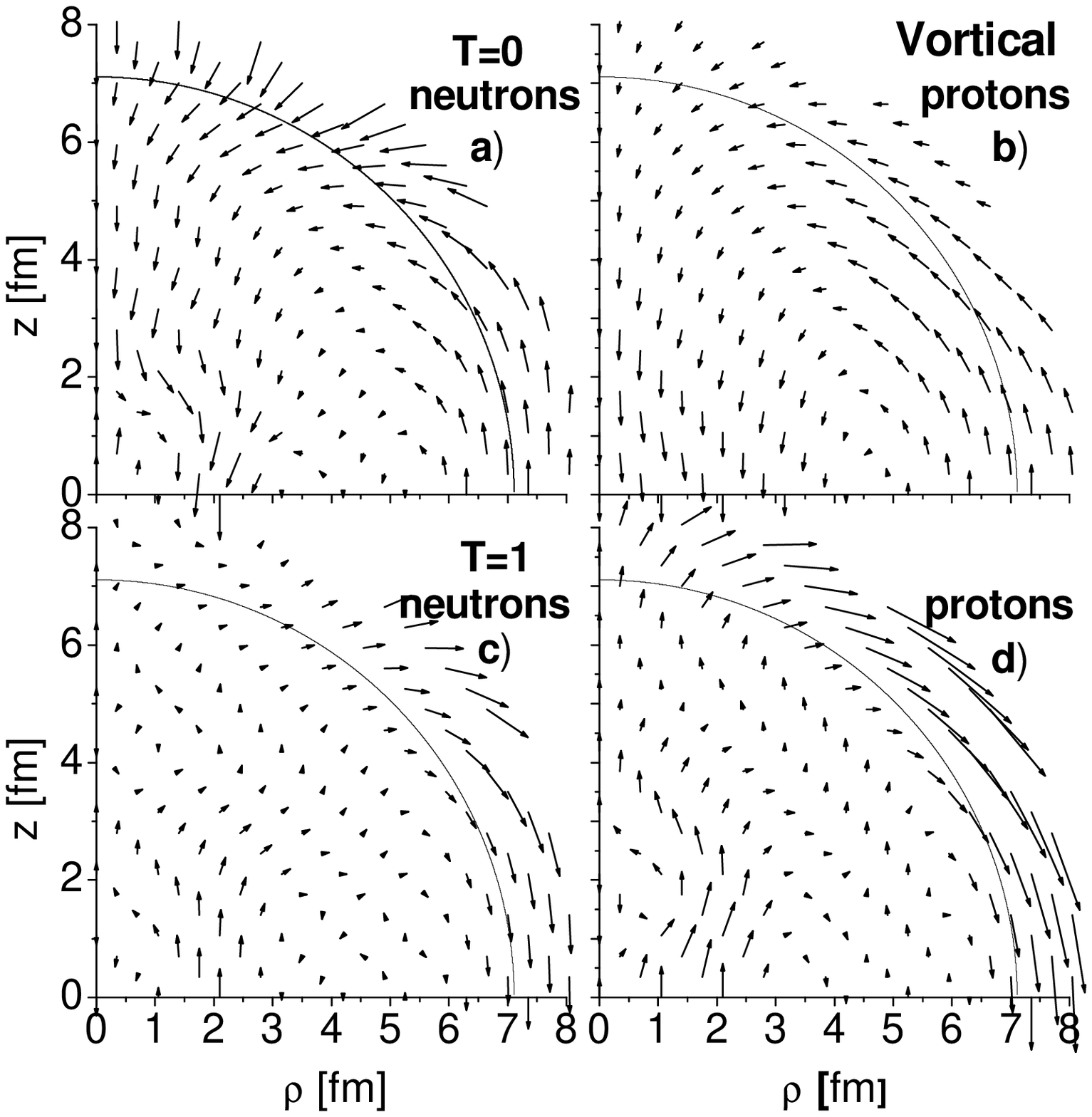}
\caption{ \label{fig:fig7}
a)-b): Neutron and proton vortical velocity fields
${\vec v}^q_{\nu}({z,\rho})$ for the state
$\omega_{\nu}$ = 8.3 MeV with a maximal T=0 vortical response.
c)-d): The same for the state $\omega_{\nu}$ = 9.1 MeV with a maximal
T=1 vortical response. For a better view, the velocities are
amplified by the factors 50 (a-b) and 20 (c-d).}
\end{figure}

\begin{figure}[t]
\includegraphics[height=7.5cm,width=7.5cm]{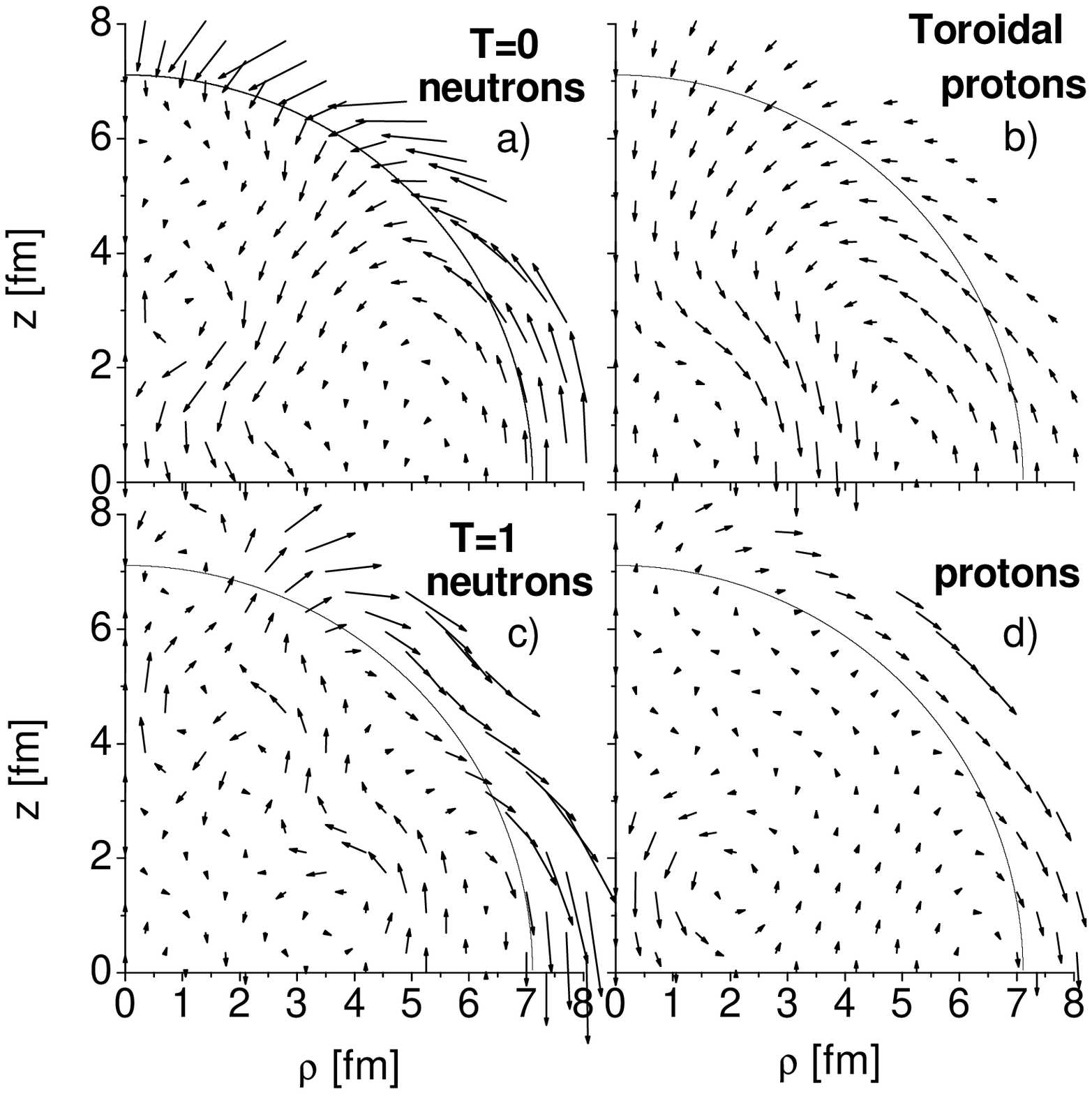}
\caption{ \label{fig:fig8}
The same as in Fig. \protect\ref{fig:fig7} but for the states $\omega_{\nu}$ = 8.7 MeV
(T=0) and $\omega_{\nu}$ = 9.8 MeV (T=1) with a maximal toroidal strength.
}
\end{figure}

\begin{figure}
\includegraphics[height=7.5cm,width=7.5cm]{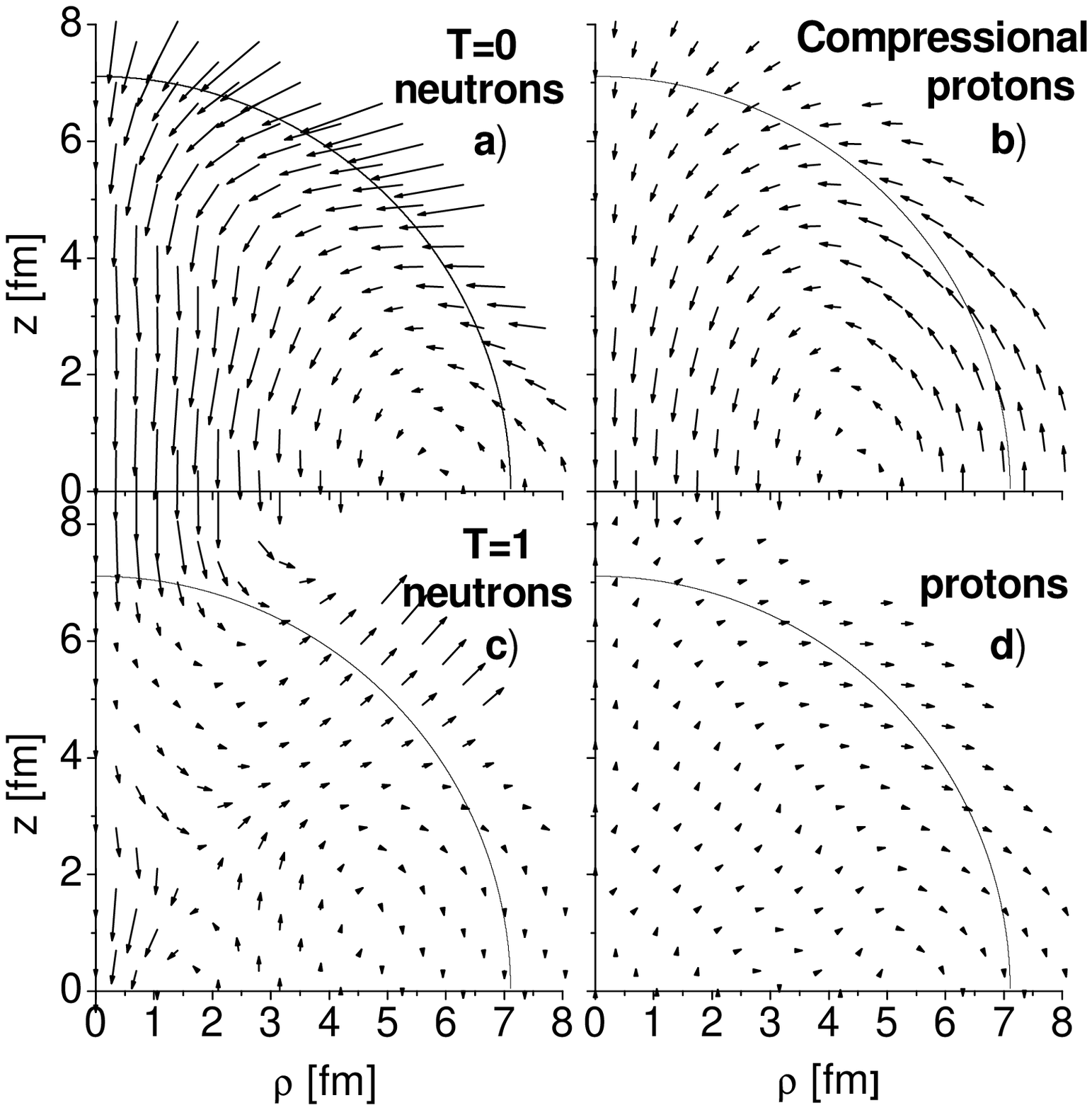}
\vspace{0.8cm}
\caption{ \label{fig:fig9}
The same as in Fig. \protect\ref{fig:fig7} but for the states $\omega_{\nu}$ = 7.1 MeV
(T=0) and $\omega_{\nu}$ = 8.8 MeV (T=1) with a maximal compression
strength in the low-energy branch.}
\end{figure}

Results of the calculations for the nucleus $^{208}$Pb are
presented in Figs. \ref{fig:fig1}-\ref{fig:fig9}.

In Figure \ref{fig:fig1}, the VM, TM, and CM strengths in T=0 and 1 channels
are compared. The strengths are computed for the transition operators
(\ref{vort_dip_oper})-(\ref{comp_dip_op}). For the CM, the current-dependent
operator $\hat{M}_{\text com}$ is used. Unlike its familiar density-dependent
counterpart $\hat{M}'_{\text com}$, it has the same dimension as the VM and TM
operators and so is more suitable for the comparison of the modes. All the
modes are  calculated with total nuclear current $j_{nuc}$.

Figure \ref{fig:fig1} shows that all the modes have basically two
broad branches, a low-energy branch (LEB) at 5-20 MeV and a high-energy
branch (HEB) at 25-40 MeV. The VM is well presented in both branches
while TM and CM are mainly localized in LEB and HEB, respectively.
Such a double-branch structure was found for the TM and CM in most of the previous
theoretical studies \cite{Pa07}. It is most probably  related to E1 transitions with
$\Delta N$=1 and 3 where $N$ is the principle shell number.

The double-branch structure of the E1(T=0) strength was confirmed by various
experiments \cite{Cl01,Morsch_PRL_80,Morsch_PRC_83,Adams_PRC_86,Davis_PRC_97,
Yo04,Uchida_PLB_03,Uchida_PRC_04}, mainly in $(\alpha, \alpha')$ scattering
at small angles. The recent results of this reaction
\cite{Uchida_PLB_03,Uchida_PRC_04} are depicted in Fig. \ref{fig:fig1}a).
The reaction is
considered as a common tool for measurements of the dipole  CM(T=0)
\cite{Harakeh_book_01}.

As seen from the panel a) of Fig. \ref{fig:fig1}, our results do not reproduce
the experimental energies and widths of the LEB and HEB. Neither of the mode
centroids
coincides with the experimental peak energies. Note that this is a common
shortcoming of almost all theoretical studies performed within various
theoretical approaches \cite{Pa07}. Namely, the theory i) underestimates by 1-2
MeV the TM-dominated LEB and overestimates by about 4 MeV the CM-dominated HEB,
ii) yields a much broader TM distribution and a too narrow CM one. The reason
of the discrepancies is still unclear. Perhaps, this is partly caused by
neglecting the coupling with complex configurations.
Figure \ref{fig:fig1} shows that the VM and TM strengths are of the
same order of magnitude in the LEB left flank for T=0 and in the whole
LEB for T=1 (the difference between T=0 and 1 cases is explained below
in the discussion of Figs.  \ref{fig:fig2} and
\ref{fig:fig3}). Perhaps, in these regions the TM is mainly vortical.
We also see that VM dominates at the right LEB flank and is significant
in HEB where the TM contribution is weak.
The CM strictly dominates in HEB and has a noticeable tail in LEB at T=0.
The latter is because of the coupling between TM and CM \cite{Semenko_81,Kv03}.
Obviously, the difference
between VM, on the one hand, and TM and CM, on the other hand, is
mainly explained by the terms $\sim \vec Y_{10\mu}(\hat{\vec r})$
which are absent in $\hat{M}_{\text vor}$ but active in
$\hat{M}_{\text tor}$ and $\hat{M}_{\text com}$. Just because of these
terms, Fig. \ref{fig:fig1} cannot be used for a direct check of the
relation (\ref{vor=tor+com}). For the same reason, the similarity of
VM and CM strengths in the HEB cannot be considered as a signature of
the CM vorticity because VM and CM represent essentially different
kinds of the motion, vortical versus irrotational. Instead, this
rather means that both kinds of motion are presented by E1 $\Delta N$=3
transitions.

The obtained results suggest that the VM(T=0) may be hopefully disentangled
from other modes in $(\alpha, \alpha')$ at the excitation energy $\sim$ 16 MeV,
where the VM(T=0) strictly dominates. For $(e,e')$, the pygmy region 7-10 MeV
seems to be most promising to observe VM and TM. In this region a dominant
contribution $j_{12}(r)$ of the nuclear current is expected.

In Figure \ref{fig:fig2}, the isoscalar VM and TM strengths, calculated with
the complete $j_{nuc}=j_c+j_m$, convection $j_c$, and magnetization $j_m$
nuclear currents, are compared.  It is seen that the $j_m$ contribution is
weak and so the T=0 VM and TM are mainly of the convection nature. This is
especially the case for the
HEB where the $j_m$ contribution, being mainly of low-energy spin-flip
character, vanishes at all. The weakness of the $j_m$ weight in T=0 channel is
naturally explained by the low values of the gyromagnetic factors $g^q_s(T=0)$,
as mentioned in Sec. \ref{sec:calc_scheme}.  Fig. \ref{fig:fig2} also exhibits
the CM. Unlike Fig. \ref{fig:fig1}, here the familiar density-dependent
operator $\hat{M}'_{\text com}$ is used.  Following (\ref{oper_com_com'}),
$\hat{M}'_{\text com}$ is less energy-weighted than $\hat{M}_{\text com}$ and
so gives a more comparable CM strength in LEB and HEB.  As discussed above, the
CM is determined by $\nabla \cdot \vec{j}_{nuc}$ and so is purely irrotational.
It has no any contribution from $j_m$ and thus is fully of convective.

In Figure \ref{fig:fig3}, the VM, TM, and CM are shown in the T=1
channel. As compared to the previous T=0 case, we see dramatic changes in
the magnitude and composition of VM and TM.
In the LEB, these modes become stronger and dominated by
the $j_m$ contribution. The reason of the changes is obvious.
The isovector spin factors, $g^{n,p}_s(T=1)=-4.70\varsigma$,
are much larger than the isoscalar ones, $g^{n,p}_s(T=0)=0.88\varsigma$.
So, the T=1 spin contribution grows about
$(g^{n,p}_s(T=1)/g^{n,p}_s(T=0))^2 \sim 29$ times. It becomes
dominant and significantly increases the total VM and TM strengths.
Note that this effect does not concern the HEB
which remains purely convective. Besides, the $j_m$ effect is
zero for the CM.

The next point to be considered is collectivity of the modes.  To this end,
Figs. \ref{fig:fig4} and \ref{fig:fig5} compare RPA and unperturbed
particle-hole (1ph) strengths. It is seen that the RPA residual interaction
noticeably down-shifts the strength for T=0 and up-shifts it for T=1. The
maximal collective shifts (defined as a difference between RPA and 1ph peaks)
take place in the CM, where they reach 1-2 MeV in LEB and 2-4 MeV in HEB.  The
HEB shift is comparable to that of the E1(T=1) giant dipole resonance (GDR)
exhibited in Fig. \ref{fig:fig6} and so is indeed very large. This indicates
that HEB modes, VM and CM, are collective. The LEB modes, for exception of a
few high peaks, are less collective.
The LEB almost coincides with the region of the unperturbed
$1ph$ dipole strength depicted in Fig. \ref{fig:fig6} and so
for the LEB the single-particle aspect is also important.
These observations are confirmed by inspection of the detailed
structure of the RPA states and agree with the
previous studies \cite{Pa07} for the high-energy CM and low-energy TM.

The next figures take a closer view at the detailed structures of a
few most collective LEB modes.  In Figs. \ref{fig:fig7}-\ref{fig:fig9}, the VM,
TM, and CM neutron and proton velocity fields (\ref{vel}) for
particular T=0 and T=1 RPA ($\mu=0$) states are shown.
The states in the region 6-10 MeV with the maximal VM, TM, and CM
responses are considered: 8.3 MeV (VM), 8.7 MeV (TM), 7.1 MeV (CM) for T=0 and
9.1 MeV (VM), 9.8 MeV (TM), 8.8 MeV (CM) for T=1. These states are easily
recognized in Figs. \ref{fig:fig2}-\ref{fig:fig5} as highest peaks at
the left side of the LEB. The states combine collectivity
and single-particle effects:
their structure is a coherent superposition of many $1ph$ components
with maximal contributions  20-40$\%$.
The velocities are plotted in cylindrical coordinates $(z,\rho)$ and,
for simplicity, only the quadrant $(z >0, \rho > 0)$ is presented.

Figures \ref{fig:fig7}-\ref{fig:fig9} show that velocity fields are
rather involved, indicating a complex structure of the RPA states. The
clear imprints of the vortex motion are seen as local
curls. Sometimes, the fields well reproduce a typical toroidal
picture, see e.g. the T=0 proton velocities for VM and CM in panels
b) of Figs. \ref{fig:fig7} and \ref{fig:fig9}. The curls for the CM
may be explained by its strong coupling to TM in LEB.  In
Fig. \ref{fig:fig9}a), the strong dipole component is also
seen. However, in most of the panels, a large impact of the
single-particle motion, e.g. of the nodal structure of the leading 1ph
components, distorts the familiar collective TM and CM flows and
considerably complicates the picture.

\section{Conclusions}

The nuclear vorticity and relevant vortical, toroidal, and compression
modes (VM, TM, and CM) were explored on general formal grounds. The
operators of all three modes were derived as second-order terms in the
long-wavelength expansion of the electrical multipole operator
and its subsidiary  counterpart built following the concept \cite{Ra87}.
For the first time, the vortical
operator was constructed and related to its toroidal and compression
counterparts. The detailed comparison of the modes and their operators
was done. It was explicitly shown that, while VM (CM) is vortical
(irrotational) by construction, the TM is of a mixed character. The
vorticity criteria from HD and arguments based on the
decoupling to the charge conservation \cite{Ra87} were inspected. It
was shown that the latter deviates from the HD definition
and may lead occasionally to misleading conclusions as, e.g., a
vorticity of the CM.

The electric dipole VM, TM, and CM were computed and investigated in
$^{208}$Pb within the self-consistent Skyrme
random-phase-approximation (RPA) approach. Unlike most of the previous
studies, both convection and magnetization (spin) parts of the nuclear
current were taken into account and both isoscalar (T=0) and isovector
(T=1) channels of the modes were analyzed. It was shown that VM and CM
have low-energy and high-energy branches  while TM mainly appears
in the low-energy branch. The CM strictly dominates in a high energy
branch. In the T=0 channel, the VM and TM are almost
completely determined by the nuclear convection current while in
the T=1 channel, their low-energy branches are strictly dominated by the
spin current. This strong isospin effect is straightforwardly
explained by low (high) values of T=0 (T=1) spin g-factors, which
drastically changes the ratio between the convective and spin
contributions of the current. The effect cannot appear in the CM since
this irrotational mode has vanishing spin contribution.

The collectivity was found strong for the high-energy VM and CM
and rather weak for the low-energy VM, TM, and CM. In the latter case,
the velocity fields of the modes are rather involved. This is partly a
consequence of the complex structure of the RPA states mixing
the collective with detailed $1ph$ contributions. This holds in particular
for the vorticity which seems to be of both single-particle and
collective origin.

The VM, TM, and CM were shown to be closely related and, at the same time,
display considerable differences in their detailed strength distribution. These
modes seem to represent one family with complementing aspects. It would be
interesting to analyze the results of our study more deeply and use them  to
disentangle the modes in the $(e,e')$ and hadron reactions. This is in our next
plans.

\section*{Acknowledgments}
The work was partly supported by the DFG RE-322/12-1, Heisenberg-Landau
(Germany - BLTP JINR), and Votruba - Blokhintsev (Czech Republic - BLTP JINR)
grants. W.K. and P.-G.R. are grateful for the BMBF support under contracts 06
DD 9052D and 06 ER 9063. Being a part of the research plan MSM 0021620859
(Ministry of Education of the Czech Republic) this work was also funded by
Czech grant agency (grant No. 202/09/0084). The work of P.V. was partly
supported by the Academy of Finland and the University
of Jyv\"askyl\"a within the FIDIPRO program.

\appendix

\section{Removal of the charge conservation constraint}
\label{Sec:_constraint}

In Sec. III A, the vortical operator
$\hat{M}_{\text vor}(E\lambda\mu, k)$, completely unconstrained by the
continuity equation, is built from the $\hat{M}(E\lambda\mu, k)$ operator by
replacing the  curl of the nuclear current $\vec{\nabla} \times
\hat{\vec{j}}_{\text nuc}(\vec{r})$ by the vortical quantity
\begin{equation}
\hat{\vec \omega}_{\lambda} = [\vec{\nabla} \times \hat{\vec{j}}_{\text nuc}(\vec{r})]_{\lambda}
- \frac{i}{\lambda}kc \; [\vec{\nabla}\hat{\rho}(\vec{r})  \times \vec{r}]_{\lambda}
\label{app_1}
\end{equation}
where the terms $[...]_{\lambda}$ assume $\lambda$-components of the multipole
expansion of the values inside the brackets.
Below we present  arguments which motivate (\ref{app_1})
and compare it with the prescription \cite{Ra87}.
For the simplicity, we consider only the convection current and
neglect isospin.

\subsection{Simple arguments}

Using the HD  definition of the velocity field (\ref{vel_curr}),
we may write the {\it truly vortical} quantity
\begin{equation}\label{curl_v}
\rho_0 (\vec r) \vec{\nabla} \times {\vec v}_{\nu}(\vec r)
=
\vec{\nabla} \times \delta {\vec{j}}_{\nu}(\vec{r})
- \vec{\nabla} \rho_0 (\vec{r}) \times \hat{\vec{v}}_{\nu}(\vec{r}) \;  .
\end{equation}
Except for the second velocity-dependent term, the r.h.s. of (\ref{curl_v})
reminds the operator construction (\ref{app_1})
and thus may be used  for justification of $\hat{\vec \omega}_{\lambda}$
as a vortical quantity.

To make (\ref{curl_v}) closer to (\ref{app_1}), it is worth to express
$\vec{r}$ in terms of the (global)
  velocity operator
\begin{equation}\label{vel_[H,r]}
\hat{\vec{\text v}} = \dot{\hat{\vec{r}}}
= \frac{i}{\hbar}[\hat{H},\hat{\vec r}]= ikc \hat{\vec r}
\quad.
\end{equation}
This suggests  the replacement
\begin{equation}\label{replacement}
\hat{\vec{v}}  \to  ikc \hat{\vec r} \;
\end{equation}
in (\ref{curl_v}) and thus casts (\ref{curl_v}) to a form similar to
(\ref{app_1}) (up to the multiplier $1/\lambda$).  Note, however, that the
velocities in (\ref{curl_v}) and (\ref{vel_[H,r]}) are not the same. The
replacement of the
  velocity field ${\vec v}_{\nu}(\vec r)$ by a global velocity becomes
  strictly valid only in the sum-rule limit when all multipole
  strength is concentrated in one collective state. It remains
  probably an acceptable step for distributed spectra which often
  gather around a strongly collective mode.

\subsection{Correspondence of the recipes}

It is also worth to relate the recipe (\ref{vtd}) for the transition
densities \cite{Ra87} with our recipe (\ref{12}) and its analog (\ref{app_1}) for
the operators. As was mentioned in  Sec. III A, both recipes serve to build
the vortical quantities, though by different ways.
While (\ref{vtd})  excludes from the current the $j_{\lambda \:\lambda-1}(r)$ terms,
the recipe (\ref{12})  leads to exact compensation of the
lowest-order $k$-terms in the vortical operator.

One may show that the recipes (\ref{vtd}) and (\ref{12})
actually correspond each other. This may be done by treating (\ref{12})
in terms of the transition densities (\ref{td}) and currents (\ref{ctd}).
Using the relation between the density and current expansion multipoles
\begin{eqnarray}\label{rho_lam}
kc  \rho_{\lambda}(r) =
&&-\sqrt{\frac{\lambda}{2\lambda +1}}
\langle \frac{d}{dr}-\frac{\lambda -1}{r} \rangle
j_{\lambda \lambda -1}(r)
\\ \nonumber
&&+\sqrt{\frac{\lambda +1 }{2\lambda +1}}
\langle \frac{d}{dr}+\frac{\lambda + 2}{r}\rangle
j_{\lambda \lambda +1}(r) \; ,
\end{eqnarray}
one may show that (\ref{12}) indeed leads to the unconstrained
vortical transition density (\ref{vtd}). In this case, the second r.h.s.
term of Eq. (\ref{app_1}) has the form
\begin{equation}\label{r_grad_rho}
ikc \; [\vec{\nabla} \delta \rho_{fi}(\vec{r}) \times  \vec{r}]
= \sum_{\lambda \mu}
a^{fi}_{\lambda \mu} S_{\lambda \lambda} (r) \:
\vec{Y}_{\lambda \lambda \mu}(\hat{\vec r})
\end{equation}
with the multipoles
\begin{equation}
\label{SLL}
S_{\lambda \lambda} (r)= \sqrt{\lambda (\lambda + 1)} kc \rho_{\lambda} (r) \; .
\end{equation}
Being scaled by $1/\lambda$, these multipoles coincide with the expansion coefficients
of $\delta \vec{S}_{fi}(\vec{r})$ in (\ref{std}).

\section{Extraction of spurious admixtures}
\label{Sec:_spurious}

The isoscalar electric dipole VM, TM, and CM may have spurious
admixtures related to the center of mass motion of the nucleus. There
are various methods to derive the center of mass
corrections (c.m.c.), see e.g. \cite{Pa07,Ri80,Li89,Gi81,Sch91}.  Most
of the methods neglect the velocity-, spin-, and spin-orbit-dependent
terms in the nuclear interaction and assume a closure relation
where all excitation is contained in one single collective state.
Hence the methods are approximate.
Nevertheless, being simple and convenient, they
are widely used in the Skyrme-Hartree-Fock calculations \cite{Pa07}.
Below we use the method \cite{Gi81} to derive the c.m.c. for isoscalar
TM and CM. For the isoscalar VM, the c.m.c. is shown to be zero.

Let's consider a system with the Hamiltonian $\hat{H} = \hat{T}+\hat{V}$
whose interaction $\hat{V}$ does not depend on spin and
velocities. Then, in the isoscalar (T=0) case, for any one-body
external field $\hat{F}=\sum_{i}^A f(\vec{r_i})$, where $f(\vec{r_i})$
is an arbitrary function of nucleon coordinates, one may write the sum
rules \cite{Suzuki_Rowe_NPA}
\begin{eqnarray}
\sum_{\nu} \omega_{\nu}
\langle 0|\hat{\rho}(\vec{r})|\nu\rangle
\langle \nu|\hat{F}|0\rangle &=&
-\frac{1}{2m} \vec{\nabla} \cdot [\rho_0(\vec{r})\vec{\nabla} f (\vec{r})]
\; ,
\label{B_rho}
\\
\sum_{\nu}
\langle 0|\hat{\vec{j}}(\vec{r})|\nu\rangle
\langle\nu|\hat{F}|0\rangle
&=& \frac{1}{2mi}\rho_0(\vec{r}) \vec{\nabla} f(\vec{r})
\label{B_j}
\end{eqnarray}
for the isoscalar transition density $\langle
0|\hat{\rho}(\vec{r})|\nu\rangle$ and current $\langle
0|\hat{\vec{j}}(\vec{r})|\nu\rangle$. The sum runs through the full
set of the excitation eigenstates $|\nu\rangle$
($\hat{H}|\nu\rangle=\omega_{\nu}|\nu\rangle)$ with the eigenvalues
$\omega_{\nu}$. Further, $\rho_0(\vec{r})=\langle 0 |\sum_{i}^A \delta
(\vec{r}-\vec{r_i})|0\rangle$ is the nuclear ground state density, and
$m$ is the nucleon mass.

If the external field mainly excites one single collective state $\bar\nu$, then
only the term ($\nu=\bar\nu$) survives in (\ref{B_rho})-(\ref{B_j})
and the corresponding transition density and current are
uniquely specified
through $\rho_0(\vec{r})$ and $f(\vec{r})$ \cite{Suzuki_Rowe_NPA}:
\begin{eqnarray}
\langle 0|\hat{\rho}(\vec{r})|\bar{\nu}\rangle
 &=&
-\frac{1}{2m} \frac{1}{\omega_{\bar{\nu}}\langle\bar{\nu}|\hat{F}|0\rangle}
\vec{\nabla} \cdot [\rho_0(\vec{r})\vec{\nabla} f (\vec{r})] \; ,
\\
\label{B_rho_c}
\langle 0|\hat{\vec{j}}(\vec{r})|\bar{\nu}\rangle
&=& \frac{1}{2mi}\frac{1}{\langle\bar{\nu}|\hat{F}|0\rangle}
\rho_0(\vec{r}) \vec{\nabla} f(\vec{r}) \; .
\label{B_j_c}
\end{eqnarray}
Hence we get simple recipes for the transition densities and currents,
\begin{eqnarray}
\delta \rho(\vec{r})
&\propto& \vec{\nabla} \cdot [\rho_0(\vec{r})\vec{\nabla} f (\vec{r})] \; ,
\label{B_rho_pro}
\\
\delta \vec{j}(\vec{r})
&\propto& \rho_0(\vec{r}) \vec{\nabla} f(\vec{r})
\propto \rho_0(\vec{r}) \vec{v}(\vec{r}) \; ,
 \label{B_j_pro}
\end{eqnarray}
to be used in further c.m.c. derivation. Here (\ref{B_j_pro})
actually defines  an irrotational flow with the velocity
$\vec{v}(\vec{r}) \sim \vec{\nabla} f(\vec{r})$.

A change of the expectation value $\langle\hat{O}\rangle$ of any
one-body operator $\hat{O}=\sum^A_i o(\vec{r}_i)$, caused by an
external field $\hat{F}$, is
\begin{eqnarray}
\delta \langle\hat{O}\rangle \:&=& \int d^3r \: \delta \rho(\vec{r})\: o(\vec{r})
\label{B5_1} \\
\quad &\propto&
\int d^3r \: o(\vec{r}) \: \vec{\nabla} \cdot [\rho_{0}(\vec{r})\: \vec{\nabla} f(\vec{r})] =
\label{B5_2} \\
\qquad
&=& - \int d^3r \: \rho_{0}(\vec{r}) \:\vec{\nabla} f(\vec{r}) \cdot \vec{\nabla} o(\vec{r})
\label{B5_3}
\\
\qquad
&\propto& - \int d^3r \: \delta \vec{j}(\vec{r}) \cdot \vec{\nabla} o(\vec{r})
\label{B5_4}
\end{eqnarray}
where we use the relations (\ref{B_rho_pro})-(\ref{B_j_pro}).
Both $\delta \rho(\vec{r})$ and $\delta \vec{j}(\vec{r})$
may be applied to determine $\delta \langle\hat{O}\rangle$. These
cases are suitable for the modes determined by the density-dependent
and current-dependent operators, respectively. Note that
Eq. (\ref{B5_4}) with $\delta \vec{j}(\vec{r})$ is general and may be derived
and used by itself, regardlessly to the formalism (\ref{B_rho})-(\ref{B5_3})
and character of the flow. The quantity $\delta \vec{j}(\vec{r})$ is then
determined by  Eq. (\ref{vel_curr}) and, unlike (\ref{B_j_pro}),
the velocity $\vec{v}(\vec{r})$ of the flow can be not only
irrotational but also vortical or mixed. Altogether, the relations
(\ref{B5_1})-(\ref{B5_4}) may be reduced to an  expression
\begin{equation}\label{B_gen}
\delta \langle \hat{O} \rangle \:\approx
\int d^3r \: \rho_{0}(\vec{r}) \:\vec{v}(\vec{r}) \cdot \vec{\nabla} o(\vec{r})
\end{equation}
with $\vec{v} (\vec{r})$ covering both irrotational (\ref{B_j_pro}) and more general cases.

If $\langle\hat{O}\rangle$ is invariant with respect to the field $\hat{F}$,
then the requirement $\delta \langle \hat{O} \rangle = 0$ is kept and
(\ref{B_gen}) gives
\begin{equation}\label{B6}
\int d^3r \: \rho_{0}(\vec{r}) \:\vec{v} (\vec{r}) \cdot \vec{\nabla} o(\vec{r})=0
\end{equation}
For
\begin{equation}\label{o_E1}
o(\vec{r}) = r Y^*_{1\mu}(\hat{\vec r}),
\end{equation}
the operator $\hat{O}=\sum^A_i o(\vec{r}_i)$
describes the c.m. coordinate and (\ref{B6}) means that
this coordinate is not affected by the field $\hat{F}$.
By using
\begin{equation}
\vec{\nabla}o(\vec{r}) = \vec{\nabla} r Y^*_{1\mu}(\hat{\vec r})
= \sqrt{3}\:\vec{Y}^*_{10\mu}(\hat{\vec r}) ,
\end{equation}
the conditions (\ref{B6}) is cast  into
\begin{equation}
\label{B9}
\int d^3r \rho_{0}(\vec{r}) \: \vec{v}(\vec{r})
\cdot \vec{Y}^*_{10\mu}(\hat{\vec r}) \: = 0 \; .
\end{equation}

The condition (\ref{B9}) may be applied to the density-dependent CM operator
$\hat M'_{\text com}(E1\mu)$ by imposing the irrotational velocity
\begin{equation}
\label{f_com}
\vec{v}_{com'}(\vec{r})=\vec{\nabla} f_{\text com}(\vec{r})
= \vec{\nabla} Y_{1\mu}(\hat{\vec r})(r^3 - \eta r)
\end{equation}
where the second term  is the actual c.m.c. with the coefficient $\eta$
to be determined from (\ref{B9}). Then, assuming spherical nuclei
($\rho_{0}(\vec{r})= \rho_{0}(r))$ and using
\begin{equation}
\int d^3r \vec{Y}^*_{\lambda l \mu}(\hat{\vec r})
\cdot \vec{Y}_{\lambda' l' \mu'}(\hat{\vec r})
= \delta_{ll'} \delta_{\lambda \lambda'} \delta_{\mu \mu'} \; ,
\label{B10}
\end{equation}
we get
\begin{equation}\label{cmc_TM}
\eta  = \frac{5}{3}\:\langle r^2\rangle_{0}
\end{equation}
with
\begin{equation}\label{cmc_VM}
\langle r^2\rangle_{0} = \frac{\int_0^\infty r^4 \rho_{0}(r) dr }{\int_0^\infty r^2 \rho_{0}(r) dr}
\end{equation}
and finally the expression (\ref{comp_oper}) for $\hat M'_{\text com}(E1\mu)$.

For the current-dependent VM, TM, and CM operators
(\ref{vort_dip_oper})-(\ref{comp_dip_op}), we put to (\ref{B9})
the velocities
\begin{eqnarray}\label{dj_vor}
\vec{v}_{\text vor}(\vec{r}) &=&  r^{2} \vec Y_{1 2 \mu}(\hat{\vec r})
- \eta \vec Y_{1 0 \mu}(\hat{\vec r})
\; ,
 \\
\label{dj_tor}
\vec{v}_{\text tor}(\vec{r}) &=&  \frac{\sqrt{2}}{5}r^2\vec Y_{12\mu}(\hat{\vec r})
+ \vec Y_{10\mu}(\hat{\vec r}) (r^2- \eta )
\; ,
\\
\label{dj_com}
\vec{v}_{\text com}(\vec{r}) &=& \frac{\sqrt{2}}{5}r^2\vec Y_{12\mu}(\hat{\vec r})
- \vec Y_{10\mu}(\hat{\vec r}) (r^2- \eta )
\; ,
\end{eqnarray}
where the second terms with $\eta$ are the relevant c.m.c..
For exception of $\vec{v}_{\text com}(\vec{r})$, these velocities are not
reduced to the gradient form. Actually, they are
taken in the form of the external fields involved in the operators
(\ref{vort_dip_oper})-(\ref{comp_dip_op})
and generating the corresponding modes. Such presentation is in accordance
with the self-consistent treatment of nuclear excitations \cite{Bo74}, which
is done here in terms of small variations $\delta \vec{j}(\vec{r})$
of the nuclear current.

By using (\ref{dj_vor})-(\ref{dj_com}), the requirement (\ref{B9}) gives
\begin{equation} \label{cmc_v}
\eta=0
\end{equation}
for VM and
\begin{equation}\label{cmc_VMeta}
\eta =  \:\langle r^2\rangle_{0}
\end{equation}
for TM and CM. Thus we get the corrected expression (\ref{tor_dip_oper})-(\ref{comp_dip_op})
for $\hat M_{\text tor}(E1\mu)$ and $\hat M_{\text com}(E1\mu)$.
Note that vector harmonics $\vec Y_{1 2 \mu}(\hat{\vec r})$ related to the vorticity
do not contribute to the c.m.c.. This reflects the physical fact that vorticity,
being a curl flow, must be fully decoupled from the c.m. translation motion.
Hence the c.m.c. is zero for the VM. On the other hand, the TM is not
completely vortical and so its c.m.c. does not vanish.

Note that the above c.m.c. are approximate. Indeed, the calculations give for the VM,
TM, and CM responses two broad structures, which actually do not meet the
sum-rule condition of excitation of a single collective state.
Besides, the prescription \cite{Suzuki_Rowe_NPA} uses the commutator
$[\hat{H}, \hat{\rho}]$ where $\hat{H}$ is assumed not to include the
terms with velocity-, spin-, and spin-orbit
dependence. However, the effect of spin-dependent terms
in the commutator is obviously zero.
The momentum (velocity)-dependent interaction does not matter for
the Galilean-invariant Skyrme functional (the most common case) but may be
important if this invariance is violated. The spin-orbit interaction
may affect the c.m.c..

\section{Nuclear density and current operators}
\label{Sec:_density_oper}

The density operator reads
\begin{equation}\label{dens_oper}
\hat{ \rho} (\vec r)= \sum_{q =n,p}
e_{\text{eff}}^q \sum_{k \epsilon q}(\delta({\vec r} - {\vec r}_k)
\end{equation}
where $e_{\text{eff}}^q$ are proton and neutron
effective charges.

The operator of the full nuclear current consists of
the convective and magnetic (spin) parts \cite{BMv1}
\begin{equation}\label{full_j}
 \hat{\vec j}_{\text nuc}(\vec r)=
 \hat{\vec j}_c(\vec r) + \hat{\vec j}_m(\vec r)
= \frac{e\hbar}{m} \sum_{q =n,p}(\hat{\vec j}_c^q(\vec r) + \hat{\vec j}_m^q(\vec r))
\end{equation}
where
\begin{eqnarray}
\hat{\vec j}^q_c(\vec r)&=& -i e_{\text{eff}}^q
\sum_{k \epsilon q}(\delta({\vec r} - {\vec r}_k) {\vec \nabla}_k
+ {\vec \nabla}_k \delta({\vec r} - {\vec r}_k)) ,
\\
\hat{\vec j}^q_m(\vec r)&=&  \frac{g^q_{s}}{2} \sum_{k \epsilon q}
{\vec \nabla} \times \hat{\vec s}_{qk} \delta({\vec r} - {\vec r}_k) ,
\end{eqnarray}
and $\hat{\vec s}_q$ is the spin operator, $\mu_N$ is the nuclear magneton,
$g^q_{s}$ is the spin g-factor, $k$ numerates the nucleons.

The T=0 modes use the values
\begin{equation}\label{gf_T0}
e_{\text{eff}}^n=e_{\text{eff}}^p=1,
\quad g^{n,p}_{s}(T=0)=\frac{1}{2}(g_{s}^n+g_{s}^p)
\end{equation}
while the T=1 modes employ
\begin{equation}\label{gf_T1}
e_{\text{eff}}^n=-e_{\text{eff}}^p=-1,
\quad  g^{n,p}_{s}(T=1)=\frac{1}{2}(g_{s}^n-g_{s}^p) \; .
\end{equation}

\section{SRPA equations and generator operators}
\label{Sec:_SRPA}

The SRPA Hamiltonian is self-consistently derived \cite{nest_PRC_02,nest_PRC_06}
from the functional
\begin{equation}\label{eq:func}
\mathcal{E} =
  \mathcal{E}_{\mathrm{kin}} + \mathcal{E}_{\mathrm{Sk}}+ \mathcal{E}_{\mathrm{pair}}
+ \mathcal{E}_{\mathrm{Coul}}
\end{equation}
involving kinetic-energy, Skyrme, pairing and Coulomb terms. The
Skyrme functional $\mathcal{E}_{\mathrm{Sk}} (\rho, \tau, \vec{J},
\vec{j}, \vec{s}, \vec{T})$ depends on time-even (nucleon $\rho$,
kinetic-energy $\tau$, spin-orbit $\vec{J}$) and time-odd (current
$\vec{j}$, spin $\vec{s}$, vector kinetic-energy $\vec{T}$) densities.
The Hamiltonian reads \cite{nest_PRC_02,nest_PRC_06}
\begin{equation}\label{eq:Hamiltonian_residual}
    \hat{H} = \hat{h}_{\mathrm{HFB}} + \hat{V}_{\mathrm{res}}
\end{equation}
where $\hat{h}_{\mathrm{HFB}}$ is the HFB mean field
\begin{equation}\label{eq:HFBmf}
    \hat{h}_{\mathrm{HFB}} = \int d^3r \sum_{\alpha_+}
    [\frac{\delta E}{\delta J_{\alpha_+}(\vec{r})}]
    \hat{J}_{\alpha_+}
\end{equation}
and $\hat{V}_{\mathrm{res}}$ is the separable  residual interaction
\begin{equation}\label{eq:residual_interaction}
    \hat{V}_{\mathrm res} = \frac{1}{2} \sum_{k, k' = 1}^{K}
    ( \kappa_{k k'}
    \hat{X}_k \hat{X}_{k'}+ \eta_{k k'} \hat{Y}_k \hat{Y}_{k'})
\end{equation}
with one-body operators
\begin{eqnarray}
    \label{eq:operators}
    \hat{X}_k &=& i \int d^3r d^3r'
    \sum_{\alpha_+,\alpha'_+}
    \frac{\delta^2 \mathcal{E}}
    {\delta J_{\alpha_+}  \delta J_{\alpha'_+}}
    \langle[\hat{P}_k , \hat{J}_{\alpha_+} ]\rangle
    \hat{J}_{\alpha'_+} ,
    \nonumber\\
\hat{Y}_k &=& i \int \int d^3r d^3r'
    \sum_{\alpha_-,\alpha'_-}
    \frac{\delta^2 \mathcal{E}}
    {\delta J_{\alpha_-} \delta J_{\alpha'_-}}
    \langle[\hat{Q}_k , \hat{J}_{\alpha_-} ]\rangle
    \hat{J}_{\alpha'_-} ,
    \nonumber
\end{eqnarray}
and inverse strength matrices
\begin{equation}
\label{eq:kappa_eta}
  \kappa_{k k'}^{-1 } =
  - i \langle [\hat{P}_{k},{\hat X}_{k'}] \rangle \; , \quad
  \eta_{k k'}^{-1 }
  = -i
  \langle [\hat{Q}_{k},{\hat Y}_{k'}] \rangle \; .
\end{equation}
Here $\alpha_+$and $\alpha_-$ enumerate time-even $J_{\alpha_{+}}$ and
time-odd $J_{\alpha_{-}}$densities, respectively; $\hat
J_{\alpha_{\pm}}$ are the density operators; $\hat{Q}_{k}$ and
$\hat{P}_{k}=i[\hat H,\hat{Q}_{k}]$ are time-even and time-odd
hermitian generator operators. The operators of the residual
interaction $\hat{X}_k$ and $\hat{Y}_k$ are time-even and time-odd by
construction, respectively.

The single-particle Hamiltonian $\hat{h}_{\mathrm{HFB}}$ is determined
by the first functional derivatives of the initial functional
(\ref{eq:func}) while operators $\hat{X}_k$ and $\hat{Y}_k$ are driven
by the second functional derivatives of the same functional. The
residual interaction includes all the possible terms arising from
(\ref{eq:func}). Hence the model is fully self-consistent.  The number
$K$ of separable terms in (\ref{eq:residual_interaction}) is
determined by the number of the generator (input) operators
$\hat{Q}_{k}$. Usually we have $K=3 - 5$. This results in a low rank of
the RPA matrix and so in an efficient calculation scheme.

The SRPA formalism itself does not prescribe the form of the
generators ${\hat Q}_{k}$ and ${\hat P}_{k}$. At the same time, their
choice is important for a fast converge of the factorized residual
interaction $\hat{V}_{\rm res}$ to the true one with a minimal number
of separable terms. The set of the generating operators is introduced
so as to initiate in the nucleus all the relevenat kinds of motion
for the considered modes. For time-even modes, the initial generators
${\hat Q}_{k}$ are chosen first and then their time-odd counterparts
are determined from $\hat{P}_{k}=i[\hat H,\hat{Q}_{k}]$.  Instead for
time-odd modes, the initial generators ${\hat P}_{k}$ are inserted and
then their time-even counterparts $\hat{Q}_{k}=i[\hat H,\hat{P}_{k}]$
are determined.  The
generators may be arbitrarily and separately scaled, which does not
influence the results.  The coupling of the modes (e.g. of electric
and magnetic ones in deformed nuclei) may request both time-even and
time-odd  generators in the set.  The optimal sets of the
generators were developed for  E1(T=1)
\cite{nest_PRC_02,nest_PRC_06} and spin-flip M1
\cite{Ve09,Nest_JPG_10,Nest_IJMPE_10} GR.

Here we use the minimal sets of the generators suitable for the description of
VM, TM, and CM. The generators cover the main parts of the corresponding
operators and take into account the coupling between the modes.
For the VM, they are
\begin{eqnarray}
\hat{P}_1 &=& \int d^3r \; \hat{\vec{j}}_{\text c}(\vec r) r^{2}
\vec Y_{1 2 \mu}(\hat{\vec r}),
\nonumber
\\
\hat{P}_2 &=& \int d^3r \; \hat{\vec{j}}_{\text m}(\vec r) r^{2}
\vec Y_{1 2 \mu}(\hat{\vec r}),
\nonumber
\\
\hat{Q}_3 &=& \int d^3r \; \rho(\vec r)
[r^3-\frac{5}{3}\langle r^2\rangle_0^{\mbox{}} r] Y_{1\mu}(\hat{\vec r}) ,
\end{eqnarray}
i.e. cover the time-odd parts of the vortical operator with the convection
and magnetization currents as well as the time-even compression operator
with the c.m.c. (to prevent generation of the spurious motion).

For the TM,  the generators read
\begin{eqnarray}\label{tor_1}
\hat{P}_1 &=& \int d^3r \; \hat{\vec{j}}_{\text c}(\vec r)
[r^2 \frac{\sqrt{2}}{5}\vec Y_{12\mu}(\hat{\vec r}) + \vec Y_{10\mu}(\hat{\vec r})
(r^2 - \langle r^2\rangle_0^{\mbox{}})] ,
\nonumber
\\ \label{tor_2}
\hat{P}_2 &=& \int d^3r \; \hat{\vec{j}}_{\text m}(\vec r)
[r^2 \frac{\sqrt{2}}{5}\vec Y_{12\mu}(\hat{\vec r}) + \vec Y_{10\mu}(\hat{\vec r})
(r^2 - \langle r^2\rangle_0^{\mbox{}})] ,
\nonumber
\\ \label{com}
\hat{Q}_3 &=& \int d^3r  \rho(\vec r)
[r^3-\frac{5}{3}\langle r^2\rangle_0^{\mbox{}} r] Y_{1\mu}(\hat{\vec r}) ,
\end{eqnarray}
i.e. cover the time-odd parts of the toroidal operator with the convection
and magnetization currents as well as the time-even compression operator.
The TM needs the c.m.c. and so now this correction is
included to all the generators.

Finally the set for the CM includes the compression operator
$\hat{Q}_3$ itself and the convective toroidal generator $\hat{P}_1$. The generators
with the magnetic current $\hat{\vec{j}}_{\text m}(\vec r)$ are not involved
since their effect on the CM is zero.

\end{document}